\documentclass[twocolumn]{autart}    % Enable this line and disable the 

\def\@authorsize{\tiny}
\usepackage{amsfonts}
\usepackage{amsmath} 
\usepackage{amssymb}
\usepackage{mathrsfs}
\usepackage{float}   
\usepackage{arydshln}      
\usepackage{xcolor}
\usepackage{anyfontsize}
\usepackage{natbib}
\setcitestyle{round}

\makeatletter
\g@addto@macro\normalsize{%
	\setlength\abovedisplayskip{9pt}
	\setlength\belowdisplayskip{9pt}
}
\makeatother

\usepackage{graphicx}          

\definecolor{ProcessCyan}{rgb}{0, 0.5, 0.68}

\usepackage[pdftex]{hyperref} % Removed 'bookmarks' duplicate option
\hypersetup{
	colorlinks=true, 
	citecolor=ProcessCyan, 
	linkcolor=ProcessCyan, 
	urlcolor=black,
	hypertexnames=false % <--- THIS IS THE FIX
}

\numberwithin{thm}{section}

\def\Elproofname{Proof.}

\edef\endfrontmatter{%
	\unexpanded\expandafter{\endfrontmatter}% the current code
	\noexpand\endNoHyper % add \endNoHyper at the end to match \NoHyper
}

\newcommand{\neweq}[1]{\mathrel{\stackrel{\makebox[0pt]{\mbox{\normalfont\tiny{#1}}}}{=}}}
\newcommand{\newgeq}[1]{\mathrel{\stackrel{\makebox[0pt]{\mbox{\normalfont\tiny{#1}}}}{\geq}}}
\newcommand{\newleq}[1]{\mathrel{\stackrel{\makebox[0pt]{\mbox{\normalfont\tiny{#1}}}}{\leq}}}

\DeclareFixedFont{\MyAuthorFont}{OT1}{cmr}{m}{n}{13pt}

\begin{document}
		
	\begin{frontmatter}
		
		\title{ \fontsize{16}{22}\selectfont An Information-Theoretic Analysis of Continuous-Time Control and Filtering Limitations by the I-MMSE Relationships\thanksref{footnoteinfo}}

		\thanks[footnoteinfo]{This paper was not presented at any meeting. Corresponding author N. Wan.}
		
		\setcounter{footnote}{-1}
		
		\thanks{\hspace{-8.5pt}${}^\dagger$ \hspace{-1pt} Authors contributed equally to this paper.}
		
		\setcounter{footnote}{-1}
		
		\thanks{\hspace{-8.5pt}${}^1$ \hspace{1pt}This paper is the full version of the article by \cite{Wan_Auto_2026} published in Automatica.}	
				
		\author[A]{Neng Wan}$^{,\dagger}$\ead{nengwan2@illinois.edu},    % Add the 
		\author[B]{Dapeng Li}$^{,\dagger}$\ead{dapeng.ustc@gmail.com},   % e-mail address 
		\author[A]{Naira Hovakimyan}\ead{nhovakim@illinois.edu}                % (ead) as shown
		
		\address[A]{University of Illinois Urbana-Champaign, Urbana, IL 61801, USA}  % Please supply                                              
		\address[B]{Southern University of Science and Technology, Shenzhen, Guangdong 518055, China}             % full addresses
		
		                             % keyword list or with the 

		\begin{abstract}                          
			While information theory has been introduced to characterize the fundamental limitations of control and filtering for a few decades, the existing information-theoretic methods are indirect and cumbersome for analyzing the limitations of continuous-time systems. To answer this challenge, we lift the information-theoretic analysis to continuous function spaces by the I-MMSE relationships. Continuous-time control and filtering systems are modeled into the additive Gaussian channels with and without feedback, and the total information rate is identified as a control and filtering trade-off metric and calculated from the estimation error of channel inputs. Fundamental constraints for this trade-off metric are first derived in a general setup and then used to capture the limitations of various control and filtering systems subject to linear and nonlinear plant models. For linear scenarios, we show that the total information rate quantifies the performance limits, such as the minimum entropy cost and the lowest achievable mean-square estimation error, in the time domain. For nonlinear systems, we provide a direct method to calculate and interpret the total information rate and its lower bound by the Stratonovich-Kushner equation. 	
		\end{abstract}
		
	\end{frontmatter}

	\section{Introduction}
	Fundamental limitations of control and filtering have been providing guidelines to the practitioners for the design of controllers and filters in a broad range of applications. In general, these limitations are rooted in the physical limitations of systems and appear as the trade-offs between the pursuit of performance and the constraints in the design. Some recognized limitations and trade-off metrics include but are not limited to the Bode-type integrals \citep{Bode_1945, Seron_1997, Fang_2017}, data rate constraints \citep{Nair_PIEEE_2007, Silva_TAC_2016}, optimum control performance \citep{Seron_TAC_1999, Tanaka_TAC_2018, Kostina_TAC_2019}, and minimum estimation error \citep{Braslavsky_Auto_1999, Charalambous_TAC_2014, Tanaka_TAC_2017}. Complex analysis and information theory are the two popular tools for investigating the fundamental limitations and trade-offs (see \citealp{Chen_ARC_2019} for a review), although their weaknesses are also very discernible. The fundamental trade-offs based on the complex analysis can hardly be applied to the time-varying and nonlinear systems \citep{Seron_1997}, and the traditional information-theoretic methods are cumbersome and indirect when dealing with the continuous-time systems \citep{Martins_TAC_2007, Li_TAC_2013}. To cope with these challenges, by resorting to the I-MMSE (mutual information -- minimum mean-squared error) relationships, a direct information-theoretic method is proposed in this paper to investigate the fundamental limitations of continuous-time linear time-invariant (LTI), linear time-varying (LTV), and nonlinear control and filtering systems without any circumvention or approximation.

	The study of fundamental limitations or trade-offs can be traced back to \cite{Bode_1945}, who proved that for continuous-time open-loop stable systems, the integral of logarithmic sensitivity over the frequency range must equal zero. \cite{Freudenberg_TAC_1985} later showed that for open-loop unstable systems, this integral equals the sum of open-loop unstable poles. Bode's integral and its variants not only characterize a control trade-off of noise attenuation on the frequency range \citep{Middleton_TAC_1991,  Wan_TAC_2020, Wan_TAC_2024}, but quantifies the optimum cost of cheap control problem over the time horizon \citep{Qiu_Auto_1993, Seron_TAC_1999}. As the dual problem of control, estimation and its performance limits also attracted a great amount of attention. \cite{Goodwin_TAC_1997} proposed a class of integral constraints, analogous to the Bode-type integrals in control, for linear filtering, prediction and smoothing systems. \cite{Braslavsky_Auto_1999} revealed the connection between Bode-type integrals and lowest achievable mean-square (MS) estimation error in the linear filtering systems under vanishing process and measurement noise. However, due to the lack of a powerful mathematical tool other than complex analysis, most of these classical results were confined in the LTI scenario (see \citealp{Seron_1997} for a compilation), with only a few exceptions trying to explore the fundamental trade-offs for time-varying and nonlinear systems \citep{Shamma_SCL_1991, Shamma_TAC_1991, Iglesias_LAA_2002}.

	As information theory was introduced to characterize the fundamental trade-offs, control and filtering systems (with more time-varying and nonlinear elements) were modeled and analyzed as communication channels and brought to the attention of researchers. \cite{Nair_SCL_2000} proved that a discrete-time LTI plant can be stabilized by quantized control if and only if the available data rate exceeds the instability rate of plant, i.e., sum of logarithmic unstable poles. \cite{Zang_SCL_2003} and \cite{Martins_TAC_2008} respectively used the input-output entropy rate difference and total information rate as an information-theoretic interpretation of discrete-time Bode's integral, and similar to the data rate constraint, these quasi-forms of Bode's integral are bounded below by the instability rate of LTI plant. Fundamental limitations of networked control systems \citep{Elia_TAC_2004, Fang_2017}, quantized control systems \citep{Tatikonda_TAC_2004a, Nair_PIEEE_2007, Silva_TAC_2016}, LQR/LQG (linear-quadratic-Gaussian) control systems \citep{Tatikonda_TAC_2004, Tanaka_TAC_2018, Kostina_TAC_2019}, stochastic switched systems \citep{Li_Auto_2013}, contractive nonlinear systems \citep{Yu_TAC_2010}, and decentralized systems \citep{Zhao_Auto_2014} have also been investigated via information-theoretic approaches, wherein the open-loop unstable poles and non-minimum phase zeros serve as a keystone in quantifying the control trade-offs. Information theory was introduced to interpret the filtering limits in the 1970’s, when entropy and mutual information were utilized to investigate the performance limits of filters \citep{Weidemann_TIT_1970, Tomita_IS_1976, Kalata_IS_1979}. More recently, the limits of entropy reduction and maximum relative entropy were used to design and analyze the Kalman filter by \cite{Delvenne_CDC_2013} and \cite{Giffin_Entropy_2014}, and rate-distortion function was utilized to describe the filtering trade-offs and synthesize filter by \cite{Charalambous_TAC_2014}, \cite{Tanaka_TAC_2017}, and \cite{Stavrou_JSTSP_2018}. Nevertheless, most of the aforementioned analyses only considered the discrete-time settings with LTI plant and relied on the Kolmogorov-Bode formula \citep{Kolmogorov_TIT_1956, Yu_TAC_2010}, Szeg{\H o} limit theorem \citep{Szego_MA_1915, Gray_2006}, and their variants \citep{Iglesias_LAA_2002, Georgiou_TAC_2017}, accompanied by the assumptions of stationary and Gaussian auto-regressive signals, to connect the entropy rate differences with either a mutual/directed information rate or a quasi-form of Bode's integral. More importantly, entropy (rate), a pivotal tool in capturing the fundamental limitations of discrete-time systems, is undefined for the continuous-time processes over the infinite-dimensional distributions (see the comments of \citealp{Kolmogorov_AMST_1963}, \citealp{Zang_SCL_2003} or \hyperref[rem24]{Remark~2.4} for a technical explanation). Therefore, although information theory is a powerful tool in analyzing the discrete-time systems, it seems inadequate for the continuous-time systems at first glance.

	To study the fundamental continuous-time trade-offs with information-theoretic tools, a feasible scheme is to discretize the original continuous-time system to a discrete-time system by sampling and then apply the established tools and methodology to the approximate system. By making this detour, \cite{Li_TAC_2013} proposed a Bode-like integral for the continuous-time LTI control systems with  causal stabilizing controllers. \cite{Fang_2017} studied the power gain from the disturbance signal to error signal, and \cite{Wan_SCL_2019} investigated the complementary sensitivity trade-offs in continuous-time LTI systems. However, the sampling-based scheme is essentially a compromise when no direct information-theoretic approach is available to the continuous-time problems; the sampling process would inevitably require multiple regularity assumptions on, for example, the time partitions and the boundedness of functions or processes \citep{Liu_Entropy_2019, Han_arxiv_2020}, and an auxiliary time-delay has to be imposed for avoiding the causality paradox \citep{Li_Auto_2013, Weissman_TIT_2013}. Another feasible scheme is to embed the input-output relations into the complex Hardy spaces, such as the $H_2$ and $H_\infty$ spaces. \cite{Braslavsky_TAC_2007} and \cite{Rojas_Auto_2008} cast the signal-to-noise ratio (SNR) constrained channel into a classical $H_2$ optimization framework and analyzed the channel capacity constraints for both continuous- and discrete-time control systems. Unfortunately, noticing that it has complex analysis and analytical functions in its core, this scheme falls short when dealing with the non-LTI systems and non-stationary signals. Moreover, with the aid of Duncan's theorem, \cite{Tanaka_CDC_2017, Tanaka_arXiv_2022} put forward a semidefinite representation of rate-distortion function to describe the trade-offs in continuous-time LTI filtering system. It is worth noting that each of the preceding information-theoretic trade-offs and methods only provides an ad-hoc interpretation for a specific type of control or filtering problem, and few result considers the non-LTI control or filtering systems.

	A natural question arising here is whether there exist a trade-off metric and a direct information-theoretic method that uniformly apply to the continuous-time control and filtering systems subject to linear and nonlinear plants. To answer this question, a unified framework based on the additive Gaussian channel model is proposed to characterize the control and filtering trade-offs, and a direct information-theoretic method relying on Duncan's theorem is put forward to compute and analyze the trade-off metrics. The main contributions of this paper are summarized as follows:

	\noindent (C1) A unified information-theoretic framework is proposed to model and capture the fundamental limitations of continuous-time control and filtering systems. An additive Gaussian channel model with or without feedback is used as a carrier to describe the control and filtering systems (\hyperref[sec23]{Sections~2.3}-\ref{sec25}). Total information rate is identified as a trade-off metric in a general setting (\hyperref[sec31]{Sections~3.1} and \ref{sec41}) and then applied to capture the fundamental limitations of continuous-time control and filtering systems subject to LTI, LTV, and nonlinear plants (\hyperref[sec32]{Sections~3.2}-\ref{sec34} and \ref{sec42}-\ref{sec44}).

	\noindent (C2) A direct information-theoretic paradigm is developed to compute and analyze the total information rate. By invoking Duncan's theorem or the continuous-time I-MMSE relationships, we lift the calculation and analysis of total information rate to infinite-dimensional function spaces (see \hyperref[prop32]{Propositions~3.2} and \ref{prop42} for a justification). In a general control and filtering setting, we show that i) total information (rate) can be directly calculated from the causal MMSE of control input in control (\hyperref[thm31]{Theorem~3.1}) or the causal MMSE of noise-free output in filtering (\hyperref[thm41]{Theorem~4.1}), and ii) total information rate is sandwiched between the feedback capacity and instability rate of plant in control (\hyperref[thm34]{Theorem~3.4}) or the channel capacity and lowest achievable MS estimation error in filtering (\hyperref[thm43]{Theorem~4.3}). Applying these results to different control and filtering systems, we then establish an equality between total information rate and the sum of open-loop unstable poles in the LTI scenario (\hyperref[prop36]{Propositions 3.6} and \ref{prop45}), and an inequality between total information rate and the sum of spectral values in the LTV scenario (\hyperref[prop311]{Propositions 3.11} and \ref{prop48}), which suggest that total information rate serves a similar role as some established trade-off metrics, such as the Bode-type integrals, entropy cost function, and minimum MS estimation error, in linear control and filtering systems. By using the Stratonovich-Kushner equation in nonlinear filtering, a direct method to calculate the total information rate in nonlinear control and filtering systems is proposed (\hyperref[prop314]{Propositions~3.14} and \ref{prop49}). No restriction is imposed on the stationarity and distributions of signals and initial states, and the auxiliary time-delay is not needed.

	The rest of this paper is organized as follows. Information theory on continuous function spaces, continuous-time additive Gaussian channel model, and continuous-time control and filtering systems are successively introduced and modeled in \hyperref[sec2]{Section~2}. Fundamental limitations of continuous-time control and filtering systems are investigated in \hyperref[sec3]{Section~3} and \hyperref[sec4]{Section~4}, respectively, and \hyperref[sec5]{Section~5} draws the conclusions.

	\noindent \textit{Notations}: $\log(\cdot)$ denotes the logarithm in base $\rm{e}$. When $A$ is a square matrix, $\textrm{tr}(A)$ and $\det(A)$ respectively denote the {trace} and {determinant} of matrix $A$. For a function $f(\tau)$, $f \in \mathbb{C}^r[0, t]$ indicates that $f(\cdot)$ is continuously $r$th order differentiable at $\tau \in [0, t]$, and $f\in\mathcal{L}^p([0, t])$ implies that the $p$-th power of $|f(\tau)|$ is Lebesgue integrable on $[0,t]$, i.e., $\int_0^t |f(\tau)|^p d\tau < \infty$. $\mathbb{C}^{0}[0, t]$ and $\mathcal{L}^2([0, t])$ will be intermittently invoked in this paper, both of which are the function spaces of infinite dimensions. For two continuous-time random processes (or signals), $x_0^t$ and $y_0^t \in \mathcal{L}^2([0, t])$, we let inner product $\langle x, y\rangle = \langle x, y\rangle^{}_{(t, \omega)}  := \int_{0}^{t}x(\tau)^\top y(\tau) d\tau$, and norm $\|x\| = \|x\|_{(t, \omega)} := \sqrt{ \langle x, x \rangle}$, whose rigorous definitions can be found in \cite{Ibragimov_1978}, \cite{Conway_1985}, and \cite{Liptser_2001}.

	\setcounter{footnote}{1}
	\section{Preliminaries and Problem Formulation}\label{sec2}
	In this section, continuous-time signals and their associated image measures are defined and employed to reinterpret some information-theoretic metrics and properties. Subsequently, an additive white Gaussian channel model, and continuous-time control and filtering systems are respectively introduced and modeled. 
	
	\subsection{Probability in Function Spaces}
	Consider a probability space $(\Omega, \mathscr{F}, \mathbb{P})$, where $\Omega$ denotes a sample space, $\sigma$-algebra (or -field) $\mathscr{F}$ represents the collection of all subsets on $\Omega$, and $\mathbb{P}$ is a probability measure. When the time parameter $t \in \mathbb{R}^+$, the $n$-dimensional continuous-time signals (or random processes) on time interval $[0, t]$ are denoted as  $x_0^t$ or $x_0^t(\omega): \Omega \rightarrow \mathcal{R}(\mathbb{R}^+, \mathbb{R}^n)$, where $\mathcal{R}(\mathbb{R}^+, \mathbb{R}^n)$ or $\mathbb{R}^{T}$, with $T$ denoting the whole or part of the positive real line, represents the space of all real functions from $\mathbb{R}^+$ to $\mathbb{R}^n$. Random variable $x_\tau$, $x(\tau)$ or $x(\tau, \omega): \Omega \rightarrow \mathbb{R}^n$ implies the value of $x_0^t$ observed or sampled at time $\tau \in [0,t]$. Discrete-time random processes $(x_{t_1}, x_{t_2}, \cdots, x_{t_k})$, which consist of finite random variables, are defined on the finite-dimensional spaces, i.e., joint distribution of $x_{t_1},  \cdots, x_{t_k}$, and continuous-time random processes $x_0^t(\omega)$, with infinitesimal time steps, are defined on the infinite-dimensional spaces. Meanwhile, we use $\mathscr{B}(\cdot)$ or $\mathscr{B}_{\cdot}$ to denote the Borel $\sigma$-algebra (or -field) of a random process or variable, and for a collection of random processes $x_0^s, 0 \leq s \leq t$, filtration $\mathscr{F}_t^x$ is defined by $\mathscr{F}_t^x := \mathscr{B}(x_0^s, 0 \leq s \leq t) \subset \mathscr{F}$. 
	
	To specify a probability measure on the continuous function space $\mathcal{R}(\mathbb{R}^+, \mathbb{R}^n)$, let $\mu_x$ denote the image (or push-forward) measure associated with random process $x_0^t$ such that $\mu_x(A) := \mathbb{P}(x_0^\tau \in A)$, where Borel set $A \subset \mathcal{R}(\mathbb{R}^+, \mathbb{R}^n)$, $x_0^\tau: \tau \rightarrow x^\tau_0(\omega)$, $\tau \in [0, t]$, and $\omega \in \Omega$. If $x_0^t$ is a standard Brownian motion (or Wiener process), $\mu_x$ is called Wiener measure. For two probability measures $\mu_x$ and $\mu_y$, the absolute continuity of $\mu_x$ w.r.t. (with respect to) $\mu_y$ is denoted by $\mu_x \ll \mu_y$. More explanations of probability measures on infinite-dimensional spaces has been given by \cite{Kuo_1975}, \cite{Maniglia_2004}, and \cite{Eldredge_arxiv_2016}.

	\subsection{Information Theory}
	With the probability measures on continuous function spaces, we now define the information-theoretic metrics for continuous-time signals. Compared with the statistical notions \citep{Cover_2012, Gamal_2011}, the following measure-theoretic notions \citep{Polyanskiy_2016, Mehta_TAC_2008} are more general and cover more varieties of signals, e.g., signals without probability density functions. Divergence, also known as Kullback-Leibler (KL) divergence and relative entropy, is a function quantifying the dissimilarity between two probability measures. 
	\begin{defn} (Divergence)\label{def21}
		For two probability measures, $\mu_x$ and $\mu_y$, the divergence $D(\mu_x \| \mu_y)$ from $\mu_x$ to $\mu_y$ is defined by
		\begin{equation*}\label{def21_eq1}
			D(\mu^{}_x \| \mu^{}_y) := \int \left(  \log \frac{d\mu^{}_x}{d\mu^{}_y} \right) d\mu^{}_x = \mathbb{E}_{\mu^{}_x}\left[ \log \frac{d\mu^{}_x}{d\mu^{}_y}  \right],
		\end{equation*}
		where $\mu_x \ll \mu_y$. In particular, if $\mu_x$ and $\mu_y$ have densities $p_x$ and $p_y$, then $D(\mu^{}_x \| \mu^{}_y) := \int \log[ p^{}_x(x) / p^{}_y(x) ] \cdot p^{}_x(x) \ dx = \mathbb{E}_{p^{}_x}[ \log{p^{}_x(x)} / {p^{}_y(x)}]$.
	\end{defn}
	When $\mu^{}_x \not\ll \mu^{}_y$, we let $D(\mu^{}_x \| \mu^{}_y) = \infty$ by convention. Differential entropy is a function that measures the amount of randomness in a random vector. 
	\begin{defn}(Differential Entropy)\label{def22}
		For a random vector $x$ subject to probability measure $\mu_x$, its differential entropy $h(x)$ is defined by
		\begin{equation}\label{def22_eq1}
			h(x) := -D(\mu^{}_x \| \mu^{}_{\rm{Leb}}) = \int \log \left( \frac{d\mu^{}_{\rm {Leb}}}{d\mu^{}_x} \right)d\mu^{}_x,
		\end{equation}
		where $\mu^{}_{\rm{Leb}}$ indicates the Lebesgue measure. In particular, if $\mu_x$ has a density $p^{}_x$, $h(x) := \int \log[1 / p^{}_x(x)] \cdot p^{}_x(x) \ dx = \mathbb{E}_{p^{}_x}[\log(1 / p^{}_x(x))]$.
	\end{defn}
	The value of $h(x)$ in \hyperref[def22]{Definition~2.2} can be positive, negative, $\pm \infty$, or undefined. When $\mu_x$ is not absolutely continuous w.r.t. $\mu_{\rm Leb}$, i.e., $\mu_x \not\ll \mu^{}_{{\rm Leb}}$, we conventionally let $h(x) = -\infty$. Mutual information is a measure of dependence between two random processes (or vectors), and mutual information rate quantifies the reliable data transmission rate of communication channel.
	\begin{defn}\label{Def23}(Mutual Information \& Mutual Information Rate)
		Mutual information $I(x_0^t; y_0^t)$ between random processes $x_0^t$ and $y_0^t$, constrained by probability measures $\mu_x$ and $\mu_y$, is defined by
		\begin{equation*}\label{def23_eq1}
			I(x_0^t;y_0^t) := \int \log \frac{d\mu_{xy}}{d(\mu_x \times \mu_y)} d\mu_{xy} ,
		\end{equation*}
		where $d\mu_{xy} / d(\mu_x \times \mu_y)$ is the Radon-Nikodym derivative. {Mutual information rate $\bar{I}(x; y)$ is then defined by
			\begin{equation*}
				\bar{I}(x; y) := \lim_{t\rightarrow\infty} \frac{I(x_0^t; y_0^t)}{t} ,  \vspace{-0.25em}
			\end{equation*} 
			when the limit exists.\footnote{{The definition of mutual information (rate) or directed information (rate) adopted in this paper is aligned with those of \cite{Pinsker_1964}, \cite{Ihara_1993}, and \cite{Weissman_TIT_2013}. When $x_0^t$ and $y_0^t$ are stationary Gaussian or jointly stationary, the limit always exists. When the signals are nonstationary, some useful discussions and calculation results have been provided by \cite{Miao_SP_2020}, \cite{Vu_NC_2009}, and \cite{Carlos_Entropy_2019}. Throughout this paper, we assume and only consider the scenario when this limit exists. Alternatively, we can define the mutual information rate as $\bar{I}(x;y):= \limsup_{t \rightarrow \infty} I(x_0^t; y_0^t) / t$, which follows the discrete-time definition of \cite{Martins_TAC_2007} and is allowed to be infinity.}}} 
	\end{defn} \vspace{-0.5em}
	Mutual information $I(x_0^t;y_0^t)$ is finite, if $\mu_{xy} \ll \mu_x \times \mu_y$. Some properties of the preceding information-theoretic quantities, such as the chain rule of mutual information, data processing inequality, symmetry and non-negativity of mutual information, invariance of entropy, and the maximum entropy, are omitted in this paper. Interested readers are referred to the work of \cite{Li_TAC_2013}, \cite{Polyanskiy_2016}, \cite{Fang_2017}, and \cite{Wan_SCL_2019} for more details. 
	\begin{rem}\label{rem24}
		There is no differential entropy (rate) for continuous-time random process $x_0^t$, since the function space $\mathcal{R}(\mathbb{R}^+, \mathbb{R}^n)$, as the domain of image measures $\mu_x$, is of infinite dimensions, while it is well-known that there is no Lebesgue measure defined on infinite-dimensional spaces. Hence, for a continuous-time random process $x_0^t$, we cannot define its differential entropy (rate) according to \hyperref[def22]{Definition~2.2}. This motivates  to investigate the continuous-time control and filtering trade-off by resorting to the divergence (rate) and mutual information (rate), in which we can replace $\mu^{}_{\rm{Leb}}$ in \eqref{def22_eq1} with more general measures, such as Wiener and Gaussian measures, that are well-defined on infinite-dimensional spaces.			
	\end{rem} \vspace{-0.3em}

	\subsection{Continuous-Time Gaussian Channel Model}\label{sec23}   \vspace{-0.3em}
	We now introduce the continuous-time Gaussian channel model, which---dating back to Shannon's seminal work \citep{Shannon_1948}—is repurposed to facilitate our investigation of control and filtering problems. Consider the following Brownian motion formulation of additive white Gaussian channel with feedback \citep{Kadota_TIT_1971, Liu_Entropy_2019}
	\begin{equation}\label{Gaussian_Channel}
		y(t) = \int_{0}^{t}\phi(\tau, y_0^\tau, m) d\tau + w(t),
	\end{equation}
	where $y(t)$ is the channel output; transmitted message $m$ is a random variable or process such that $m\in\mathscr{B}_m$; $\phi(s, y_0^\tau, m)$ is the channel input process and coding function of $m$; $y_0^\tau$ is the sample path of $y(s)$ on $s \in [0, \tau]$, and channel noise $w(t)$, independent from message $m$, is a standard Brownian motion. To describe the Gaussian channel without feedback, {which can be seen as a special case of the feedback channel}, we only need to replace the input process $\phi(\tau, y_0^\tau, m)$ with $\phi(\tau, m)$. Throughout this paper, we also invoke \eqref{Gaussian_Channel} in the differential form, which is $dy^{}_t = \phi(t, y^{t}_0, m)dt + dw^{}_t$. Some mild assumptions and constraints are imposed on the Gaussian channel~\eqref{Gaussian_Channel}:
	
	\noindent (A1)\label{ass1} Random process $y_0^t$ is measurable on $\mathscr{B}(\mathbb{R}^+) \times \mathscr{B}_m \times \mathscr{F}_t^w$. Deterministic function $\phi(\tau, y_0^\tau, m)$ is measurable on $\mathscr{F}_\tau^y \times \mathscr{B}_m$ and is defined for all adapted processes $y_0^\tau$.
	
	\noindent (A2)\label{ass2} There exists a set $B_m \in \mathscr{B}_m$ with $\mathbb{P}(B_m) = 1$ such that for any adapted process $y_0^t$ and $m \in B_m$, if $\int_{0}^t y^\top(\tau)  y(\tau) d\tau < \infty$, then $\int_{0}^{t} \phi^\top(\tau, y_0^\tau, m)   \phi(\tau, y_0^\tau, m)d\tau \break < \infty$. Meanwhile, the expectation $\mathbb{E}[ \int_{0}^{t} \phi^\top(\tau, y_0^\tau, m)  \cdot \phi(\tau, y_0^\tau, m)d\tau ]  < \infty$.
	
	\noindent (A3)\label{ass3} There exists a unique solution $y_0^t$ to the stochastic functional~\eqref{Gaussian_Channel}.
	
	\begin{rem}
		(\hyperref[ass1]{A1}) and (\hyperref[ass3]{A3}) are fundamental and commonly posited in the investigations of stochastic control and filtering, e.g., \cite{Sivan_1972}, \cite{Liptser_2001}, and \cite{Han_TIT_2016}. (\hyperref[ass2]{A2}) is often adopted as a finite-energy constraint and can be automatically fulfilled when signals are restricted in $\mathcal{L}^2([0, t])$ space. 
	\end{rem}

	Since entropy (rate), which is a pivotal tool for deriving total information (rate) $I(y_0^t; m)$ and will be investigated later as a fundamental limitation of control and filtering, is undefined for continuous-time signals, in this paper we resort to the following Duncan's theorem \citep{Duncan_SIAMJAM_1970, Kadota_TIT_1971} to directly calculate $I(y_0^t; m)$ from the causal MMSE of the channel input.	
	\begin{lem}\label{lem26}
		For the Gaussian channel~\eqref{Gaussian_Channel} subject to (\hyperref[ass1]{A1})-(\hyperref[ass3]{A3}), the mutual information $I(y_0^t; m)$ between the channel output $y_0^t$ and message $m$, is given by
		\begin{equation*}
			I(y_0^t; m) = \frac{1}{2}   \mathbb{E}[  \|  \phi\|^2 - \|  \hat{\phi} \|^2  ] = \frac{1}{2}  \mathbb{E}[\| \phi - \hat{\phi}  \|^2],
		\end{equation*}
		where $\hat{\phi}(\tau, y_0^\tau, m) = \mathbb{E}\left[ \phi(\tau, y_0^\tau, m) | y_0^\tau  \right]$.\footnote{For a function $f(\tau, y_0^\tau, m) \in \mathcal{L}^2([0, t])$,  $\|f\|^2 = \|f\|^2_{(t, \omega)} := \int_{0}^{t}f^\top(\tau, y_0^\tau, m)  f(\tau, y_0^\tau, m)  d\tau$. The notation of conditional expectation $\mathbb{E}\left[ f(\tau, y_0^\tau, m) | y_0^\tau \right]$ is the same as $\mathbb{E}\left[f(\tau, y_0^\tau, m) | \mathcal{F}^y_\tau \right]$ in the $\sigma$-algebra language.}
	\end{lem}
	
	\noindent Additionally, we can also express the total information $I(y_0^t; m)$ in terms of divergence difference by using the following lemma \citep{Eddy_AAP_2007}.	
	\begin{lem}\label{lem27}
		For the Gaussian channel~\eqref{Gaussian_Channel} subject to (\hyperref[ass1]{A1})-(\hyperref[ass3]{A3}), we have the following identities: 
		\begin{equation*}
			\begin{split}
				D(\mu^{}_{y|m}\|\mu^{}_w) & =  \mathbb{E} [  \| \phi \|^2 / 2 ]\\
				D(\mu^{}_y \| \mu^{}_w) & =   \mathbb{E}[\|\hat{\phi}\|^2 / 2],
			\end{split}
		\end{equation*}
		where $\mu^{}_w$ denotes the Wiener measure induced by the channel noise $w_0^t$, and $\hat{\phi}(\tau, y_0^\tau, m) = \mathbb{E}\left[ \phi(\tau, y_0^\tau, m) | y_0^\tau  \right]$.
	\end{lem}
	\begin{rem}
		Duncan's theorem or \hyperref[lem26]{Lemma 2.6} not only connects information theory with estimation, but provides a direct approach to compute mutual (or total) information on continuous function spaces, as the divergences or expected norms in \hyperref[lem27]{Lemma~2.7} are now defined w.r.t. Wiener measure instead of the Lebesgue measure in differential entropy. Meanwhile, \hyperref[lem26]{Lemmas 2.6} and \ref{lem27} also apply to the Gaussian channel without feedback and always hold regardless of the stationarity and distributions of the channel input and transmitted message. 
	\end{rem}

	In addition to total information $I(y_0^t; m)$, directed information $I(\phi_0^t \rightarrow y_0^t)$, as a generalization of mutual information $I(\phi_0^t; y_0^t)$ to random objects obeying causal relations \citep{Massey_ISIT_1990}, can also be used to measure the information flow from $\phi_0^t$ to $y_0^t$ and capture the fundamental limitations of control and filtering systems \citep{Silva_TAC_2016, Tanaka_CDC_2017, Kostina_TAC_2019}. For the Gaussian channel with or without feedback as \eqref{Gaussian_Channel}, total information and directed information satisfy \citep{Weissman_TIT_2013}
	\begin{equation*}
		I(\phi_0^t \rightarrow y_0^t) = I(y_0^t; m).
	\end{equation*}
	For the non-feedback channel, we also have $I(\phi_0^t \rightarrow y_0^t) = I(\phi_0^t; y_0^t)$. Directed information rate $\bar{I}(\phi_0^t \rightarrow y_0^t) := \lim_{t \rightarrow \infty} {I}(\phi_0^t \rightarrow y_0^t) /t$ is similarly defined as the mutual information rate in \hyperref[Def23]{Definition~2.3}. However, since $I(\phi_0^t \rightarrow y_0^t)$ cannot be decomposed when message $m$ consists of multiple sources, for the convenience of later analysis, we will stick with the total information (rate) in this paper. In the following, control and filtering systems are respectively modeled as an additive Gaussian channel described by \eqref{Gaussian_Channel}.

	\subsection{Continuous-Time Control Systems}\label{sec24}  \vspace{-0.25em}
	We first consider the following continuous-time control system,
	\vspace{-1.5em}
	\begin{figure}[H]
		\centering
		\includegraphics[width=0.41\textwidth]{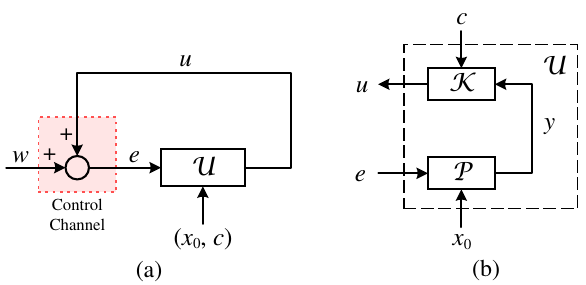} \vspace{-1em}
		\caption{Control configuration. (a) Block diagram of control systems. (b)~Lumped control process $\mathcal{U}$.}\label{fig1}
	\end{figure} 
	\noindent where $w_{(i)} \in \mathbb{C}^0[0, t]$ is a Brownian motion representing noise; $u_{(i)} \in \mathcal{L}^2([0, t])$ stands for the control signal; $e_{(i)}$ represents the error signal determined by $w_{(i)}$ and $u_{(i)}$, and subscript $(i)$ or $(ij)$ denotes the $i$-th or $ij$-th component in a vector or matrix. The lumped control mapping $\mathcal{U}$ in \hyperref[fig1]{Fig.~1(a)} can be decomposed into a plant model $\mathcal{P}$ and a control mapping $\mathcal{K}$, as \hyperref[fig1]{Fig.~1(b)} shows, in which $y_{(i)}$ is the output of plant model $\mathcal{P}$, and $c_{(i)}$, independent from $x_0$ and $w_0^t$, denotes the command signal or exogenous noise of controller $\mathcal{K}$. The notation $y_0^t$ throughout this paper has slightly different definitions in control and filtering, which will be explicitly stated before the usage. We use the following It\^o process and physical observation model to describe plant $\mathcal{P}$, which not only is general enough to cover most of real-world systems but automatically fulfills (\hyperref[ass1]{A1})-(\hyperref[ass3]{A3}),
	\begin{equation}\label{Gen_Ctrl_Dyn}
		\begin{split}
			%		dy = f(t, y, x_0)  dt + b(t)   e dt
			dx_t & = f(t, x_t)  dt + b(t, x_t)  de_t\\
			y_t & = h(t, x_t),
		\end{split}
	\end{equation}
	where $x_t \in \mathbb{R}^n$ is the internal state of $\mathcal{P}$; $f(t, x_t)$ and $b(t, x_t)$ are adapted processes, i.e., $\mathcal{F}_t$-measurable, such that $\int_{0}^{t}|f_{(i)}(\tau, y_\tau)| d\tau < \infty$ and $\int_{0}^{t} b^2_{(ij)}(\tau, y_\tau) d\tau < \infty$, and $h(t, x_t)$ is a Lebesgue measurable function. The control mapping $\mathcal{K}$ is realized by a square-integrable function
	\begin{equation}\label{Gen_Ctrl_Ctrl}
		u^{}_t = g(t, y_0^t, c_0^t),
	\end{equation} 
	with memory, or a square-integrable function $u_t = g(t, y^{}_t, c^{}_t)$ without memory. We can then collectively formulate the lumped control mapping $\mathcal{U}$ or control channel in \hyperref[fig1]{Fig.~1} as
	\begin{equation}\label{Gen_Ctrl_Err}
		de_t = u(t, e_0^t, c_0^t) dt + dw_t,
	\end{equation}
	where the lumped control process or control signal $u_t = u(t, e_0^t, c_0^t) = u_x(t, x_0^t)$ can be formulated as a function subject to the arguments $(e_0^t, c_0^t, x^{}_0)$ or $x_0^t$, depending on the statistics of interest. To align with the expression of Gaussian channel in \eqref{Gaussian_Channel}, we can rewrite \eqref{Gen_Ctrl_Err} in an integral form as $e(t)= \int_{0}^{t}u(\tau, e_0^\tau, c_0^\tau, x_0) d\tau + w(t)$, where $x_0 \in \mathbb{R}^n$ denotes the initial states of plant $\mathcal{P}$, and the following stability condition is imposed on the closed-loop system. 	
	
	\noindent (A4) The feedback system realized by \eqref{Gen_Ctrl_Dyn}-\eqref{Gen_Ctrl_Err} is internally mean-square stable, i.e., $\sup_{t\geq 0}\mathbb{E}[x^\top(t)x(t)] < \infty$.

	\subsection{Continuous-Time Filtering Systems}\label{sec25}
	To investigate the fundamental limitations of filtering, consider the filtering setup as follows
	\vspace{-0.5em}
	\begin{figure}[H]
		\centering
		\includegraphics[width=0.36\textwidth]{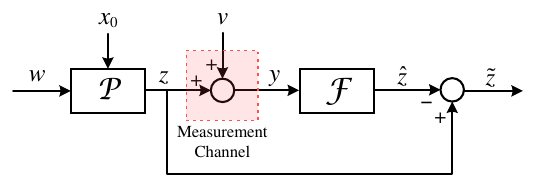} \vspace{-0.5em}
		\caption{Filtering configuration.}\label{fig2} 
	\end{figure}
	\noindent where ${w}$ represents the process input or noise; $\mathcal{P}$ denotes the plant with or without feedback; $x_0 \in \mathbb{R}^n$ is the initial state of $\mathcal{P}$; $z_{(i)} \in \mathcal{L}^2([0, t])$ is the noise-free output to be estimated; $v$ denotes the measurement noise; $y$ implies the measured output; $\mathcal{F}$ represents the filtering process; $\hat{z}$ stands for the estimate of $z$, and $\tilde{z} = z - \hat{z}$ is the estimation error. We use the following It\^o process and mathematical observation model to describe the plant $\mathcal{P}$ and measurement channel in \hyperref[fig2]{Fig.~2}:
	\begin{equation}\label{Gen_Filter_Model}
		\begin{split}
			dx_t & = f(t, x_t) dt + b(t, x_t)  [ u(t, x_t) dt +  \Sigma^{1/2}_{\bar{w}}(t) d\bar{w}_t]\\
			dy_t & = z_x(t, x_0^t)dt + \Sigma^{1/2}_v(t) dv_t,
		\end{split}
	\end{equation}
	where $x_t \in \mathbb{R}^n$ is the internal state of plant $\mathcal{P}$; $b(t, x_t)$ denotes the control matrix; $u(t, x_t)$ is the control signal if plant $\mathcal{P}$ is closed-loop; noise-free output or measurement process $z_t = z_x(t, x_0^t) = z_{\bar{w}}(t, \bar{w}_0^t, x_0)$ is Lebesgue measurable and can be expressed as a function subject to arguments $x_0^t$ or $(\bar{w}_0^t, x_0)$ depending on the statistics of interest; $d\bar{w}_t$ and $dv_t$ are vectors of Brownian components with zero drift and unit-variance rate, and positive semi-definite matrices, $\Sigma_{\bar{w}}$ and $\Sigma_v$, respectively denote the covariance matrices of $d\bar{w}_t$ and $dv_t$. Without loss of generality, in this paper we mainly consider the scenario when $\Sigma_v = I$ and $\Sigma_{\bar{w}} = \varepsilon I \geq 0$, which allows us to adjust the level of process noise in later analysis, and the filtering model \eqref{Gen_Filter_Model} can therefore be simplified as
	\begin{equation}\label{Filter_Model}
		\begin{split}
			dx_t & = f(t, x_t) dt + b(t, x_t)[u(t, x_t) dt + dw_t]\\
			dy_t & = z(t, w_0^t, x_0) dt + dv_t,
		\end{split}
	\end{equation}
	where  $dw_t = \Sigma^{1/2}_{\bar{w}} d\bar{w}_t = \sqrt{\varepsilon}I \cdot d\bar{w}_t$ is the process input, and  $z_t = z(t, w_0^t, x_0) = z_{\bar{w}}(t, \bar{w}_0^t, x_0)$ is the noise-free output. To align with the expression of Gaussian channel in \eqref{Gaussian_Channel}, we can rewrite the observation formula in \eqref{Filter_Model} as $y_t = \int_{0}^{t}z(\tau, w_0^\tau, x_0) d\tau + v_t$. Realization and formulas of the filtering process $\mathcal{F}$ can be varied depending on the design characteristics, e.g., MMSE, maximum a posteriori (MAP), minimum conditional inaccuracy. In this paper, we mainly focus on the scenario and filters, such as Kalman-Bucy filter, that minimize MS estimation error. More detailed interpretations and formulas of these optimal filters will be given later. Up until now, we have modeled the continuous-time control and filtering systems as an additive Gaussian channel with or without feedback in \eqref{Gen_Ctrl_Err} and \eqref{Filter_Model}, where fundamental limitations of control and filtering will be defined and analyzed in the following sections, respectively.

	\vspace{-0.25em}
	
	\section{Fundamental Limitations of Control}\label{sec3}
	
	\vspace{-0.25em}
	
	In this section, total information rate $\bar{I}(e; x_0, c)$ or directed information rate $\bar{I}(u \rightarrow e)$, which quantifies the reliable data transmission rate of the control channel in \hyperref[fig1]{Fig. 1}, is identified as a control trade-off metric in the general setting depicted by \eqref{Gen_Ctrl_Dyn}-\eqref{Gen_Ctrl_Err}, and then applied to capture the fundamental limitations of control systems subject to LTI, LTV, and nonlinear plants, respectively.

	\vspace{-0.5em}

	\subsection{General Control Trade-offs}\label{sec31} 
	
	\vspace{-0.5em}
	
	Based on Duncan's theorem, the following theorem provides a direct method to calculate $I(e_0^t; x_0, c_0^t)$ from the causal MS estimation error of control input.
	
	\vspace{-0.5em}
	
	\begin{thm}\label{thm31}
			For the closed-loop system described by (\ref{Gen_Ctrl_Dyn})-(\ref{Gen_Ctrl_Err}) and \hyperref[fig1]{Fig. 1}, the total information $I(e_0^t; x_0, c_0^t)$ satisfies 
			\begin{equation}\label{thm31_eq1}
				I(e_0^t; x_0, c_0^t)  = \frac{1}{2}  \mathbb{E}\left[  \| u - \hat{u}\|^2   \right]= \frac{1}{2}  \left( \mathbb{E} [ \| u\|^2 ] - \mathbb{E}[\|\hat{u}\|^2]  \right),
			\end{equation}
			where $\hat{u}(\tau, e_0^{\tau}, c_0^\tau, x_0) =  \mathbb{E}[u(\tau, e_0^{\tau}, c_0^\tau, x_0)|e_0^\tau]$, and the total information rate is given by 
			\begin{equation}\label{thm31_eq2}
				\bar{I}(e; x_0, c)  = \lim_{t\rightarrow\infty} I(e_0^t; x^{}_0, c_0^t)/t .
			\end{equation}
	\end{thm}
	
	 \vspace{-2.5em}
	 
	\begin{pf}		
		Since the control channel \eqref{Gen_Ctrl_Err} is a noisy communication channel with feedback, by applying the feedback version of \hyperref[lem26]{Lemma 2.6} to \eqref{Gen_Ctrl_Err}, we obtain the identities in \eqref{thm31_eq1}. The total information rate $\bar{I}(e; x_0, c)$ in \eqref{thm31_eq2} follows \hyperref[Def23]{Definition 2.3}, and the existence of $\bar{I}(e; x_0, c)$ is guaranteed by the boundedness of $\mathbb{E}[\|u\|^2]$ in \eqref{thm31_eq1} and (\hyperref[ass2]{A2}). \hspace*{\fill}$\blacksquare$
	\end{pf} 
	
	\vspace{-1.5em}
	
	\noindent By using \hyperref[lem27]{Lemma 2.7} and the chain rule of mutual information, we connect $I(e_0^t; x^{}_0,  c_0^t)$ with $I(e_0^t; x_0)$, the instability rate of plant $\mathcal{P}$, and express $I(e_0^t; x^{}_0,  c_0^t)$ as a difference between the divergences w.r.t. Wiener measure.
	
	\vspace{-0.5em}
	
	\begin{prop}\label{prop32}
		For the closed-loop system subject to (\ref{Gen_Ctrl_Dyn})-(\ref{Gen_Ctrl_Err}) and \hyperref[fig1]{Fig. 1}, the total information $I(e_0^t; x^{}_0,  c_0^t)$ satisfies \vspace{-0.5em}
		\begin{equation}\label{cor32_eq1}
			I(e_0^t; x^{}_0,  c_0^t) = I(e_0^t; x_0) + I(e_0^t; c_0^t|x_0),
		\end{equation}
		and can be expressed as a divergence difference,
		\begin{equation}\label{cor32_eq2}
			I(e_0^t; x^{}_0,  c_0^t) = D(\mu^{}_{e|(x^{}_0, c)}\|\mu^{}_w) - D(\mu^{}_e \| \mu^{}_w),
		\end{equation}
		where $D(\mu^{}_{e|(x^{}_0, c)}\|\mu^{}_w) = \mathbb{E}[\|u\|^2 / 2]$ and $D(\mu^{}_{e}\|\mu^{}_w) =  \mathbb{E}[\|\hat{u}\|^2 / 2]$.
	\end{prop}  
	
	\vspace{-1.5em}
	
	\begin{pf}
		\eqref{cor32_eq1} can be proved by applying the chain rule of mutual information\hspace{-1pt}\footnotemark to $I(e_0^t; x^{}_0,  c_0^t)$. \eqref{cor32_eq2} can be straightforwardly verified by replacing the divergences and mutual information with their definition forms in \hyperref[def21]{Definition 2.1} and $I(x_0^t; y_0^t) = \mathbb{E}_y[\log(d\mu_{y|x} /  d\mu_y)]$ in \hyperref[Def23]{Definition 2.3}. Meanwhile, by applying \hyperref[lem27]{Lemma 2.7} to \eqref{Gen_Ctrl_Err}, we have $D(\mu_{e|(x^{}_0, c)}\|\mu^{}_w) = \mathbb{E}[\|u\|^2 / 2]$ and $D(\mu_{e}\|\mu^{}_w) = \mathbb{E}[\|\hat{u}\|^2 / 2]$.  \hspace*{\fill}$\blacksquare$
	\end{pf} 
	
	\footnotetext{Given random vectors (or processes) $x$, $y$ and $z$, the mutual information satisfies $I(x;y, z) = I(x; z) + I(x;y|z)$, where $I(x;y|z)$ is the conditional mutual information between random vectors (or processes) $x$ and $y$ given $z$.}
	
	\vspace{-1em}
		
	\begin{rem}
		Compared with the core equation (11) of \cite{Martins_TAC_2008} and (9.117) of \cite{Cover_2012}, which express the total information as entropy differences in discrete-time setting, equations \eqref{thm31_eq1} and \eqref{cor32_eq2}, which resort to the MS estimation error and divergence difference, provides the continuous-time counterparts to calculate the total information (rate) without any discretization or approximation. 
	\end{rem}

	Before we present the trade-off properties of $\bar{I}(e; x_0, c)$, the definitions of feedback capacity and channel capacity are first given. For the feedback channel described in \eqref{Gen_Ctrl_Err} and \hyperref[fig1]{Fig.~1}, we define the ({\it information}) {\it feedback capacity}\footnote{In this paper, we adopt the mutual information version of feedback capacity following \cite{Kadota_TIT_1971a} and \cite{Ihara_1993}. In comparison, the definition of (operational) feedback capacity from a coding-theorem perspective can be found in the work of \cite{Kim_TIT_2006} and \cite{Weissman_TIT_2013}.} $\mathcal{C}_f := \sup_{(x_0, c, u)} \bar{I}(e; x_0, c)$ as the supremum of total information rate over all admissible pairs of $(x_0, c_0^t)$ and $u_0^t$. For the same channel without feedback, when the control input is subject to the power constraint, $\mathbb{E}[u_\tau^\top u^{}_\tau] \break \leq \rho(\tau), \forall \tau \in [0, t]$, we define the {\it channel capacity} $\mathcal{C}:= \sup_{\mathbb{E}[u_\tau^\top u^{}_\tau] \leq \rho(\tau)} \bar{I}(e; u)$ as the supremum of input-output mutual information rate $\bar{I}(e; u)$. The following theorem derived from \hyperref[prop32]{Proposition} \ref{prop32} shows that the information rate $\bar{I}(e; x_0, c)$ must be sandwiched between the feedback capacity $\mathcal{C}_f$ and the mutual information rate $\bar{I}(e; x_0)$. 
	\begin{thm}\label{thm34}
		For the control system described by (\ref{Gen_Ctrl_Dyn})-(\ref{Gen_Ctrl_Err}) and \hyperref[fig1]{Fig. 1}, the total information rate $\bar{I}(e; x_0, c)$ is bounded by  
		\begin{equation}\label{prop34_eq1}
			\mathcal{C}_f \geq \bar{I}(e; x_0, c) \geq \bar{I}(e; x_0),
		\end{equation}
		where $\bar{I}(e; x_0, c) = \bar{I}(e; x_0)$ if $e_0^t$ and $c_0^t$ are conditionally independent given $x_0$, {i.e.}, $c_0^t - x_0 - e_0^t$ forms a Markov chain. When the control input $u_0^t$ is further constrained by $\mathbb{E}[u_\tau^\top u^{}_\tau] \leq \rho(\tau), \forall \tau \in [0, t]$, the total information rate $\bar{I}(e; x_0, c)$ satisfies 
		\begin{equation}\label{cor36_eq1}
			\mathcal{C}_f \geq \bar{I}(e; x_0, c) + \bar{D}(\mu_e \| \mu_w) \geq \bar{I}(e; x_0) + \bar{D}(\mu_e \| \mu_w),
		\end{equation}
		where $\mathcal{C}_f = \mathcal{C} = (2t)^{-1}  \int_{0}^{t}\rho(\tau) d\tau$, {and the divergence rate $\bar{D}(\mu_e \| \mu_w) := \lim_{t \rightarrow \infty} {D}(\mu_e \| \mu_w)  / t$.}
	\end{thm} \vspace{-1em}
	\begin{pf}
		The first inequality in \eqref{prop34_eq1} can be readily obtained from the definition of feedback capacity $\mathcal{C}_f$. By the chain rule and non-negativity of mutual information, we have $I(e_0^t; x^{}_0,  c_0^t) \geq I(e_0^t; x_0)$, which implies the second inequality in \eqref{prop34_eq1}. Meanwhile, according to the chain rule, $\bar{I}(e; x_0, c) = \bar{I}(e; x_0)$ holds when $I(e_0^t; c_0^t|x_0) = 0$, which requires $e_0^t$ and $c_0^t$ be conditionally independent given $x^{}_0$, {i.e.}, $c_0^t-x^{}_0-e_0^t$ forms a Markov chain. 
		
		{When we impose a power constraint on the control input, since $D(\mu_{e|(x_0, c)}\|\mu_w) =\mathbb{E}[\|u\|^2 / 2]$ by \hyperref[prop32]{Proposition 3.2}, the feedback and channel capacities of \eqref{Gen_Ctrl_Err} then satisfy (Theorem~6.4.1 of \citealp{Ihara_1993}, and Section~4 of \citealp{Liu_Entropy_2019}):
			\begin{equation}\label{cor36_eq2}
				\mathcal{C}_f = \frac{\int_{0}^{t} \rho(\tau) d\tau}{2t} \geq \lim_{t \rightarrow \infty} \frac{\mathbb{E}[\|u\|^2]}{2  t} = \bar{D}(\mu_{e|(x^{}_0, c)}\|\mu^{}_w),
			\end{equation}
			where $\mathcal{C}_f = \mathcal{C}$; Fubini's theorem is invoked to swap the order of integration and expectation when deriving the inequality in \eqref{cor36_eq2}, and  divergence rate is given by $\bar{D}(\mu_{e|(x^{}_0, c)}\|\mu^{}_w) :=  \lim_{t \rightarrow \infty} {D}(\mu_{e|(x^{}_0, c)}\|\mu^{}_w) / t$. Dividing both sides of \eqref{cor32_eq2} by $t$, taking the limit as $t\rightarrow\infty$, and plugging \eqref{cor36_eq2} into the result give the first inequality in \eqref{cor36_eq1}, and the second inequality follows from \eqref{prop34_eq1}.}  \hspace*{\fill}$\blacksquare$
	\end{pf} \vspace{-1em}
	\noindent An immediate or information-theoretic disclosure from \hyperref[thm34]{Theorem 3.4} is that for an internally mean-square stable channel depicted by (\ref{Gen_Ctrl_Dyn})-(\ref{Gen_Ctrl_Err}) and \hyperref[fig1]{Fig. 1}, the coding scheme (or encoder/decoder) of \eqref{Gen_Ctrl_Err} must be designed such that the feedback capacity $\mathcal{C}_f$ is large enough to compensate the cost of transmitting message, gauged by $\bar{I}(e; x_0, c)$ and $\bar{I}(e; x_0)$, and the cost of attenuating channel noise, quantified by the divergence rate $\bar{D}(\mu_e \| \mu_w)$. Meanwhile, to achieve the channel capacity $\mathcal{C} = \mathcal{C}_f = (2t)^{-1}\int_{0}^{t}\rho(\tau) d\tau$, we must have $\bar{D}(\mu_e \| \mu_w) = 0$ or $D(\mu_e \| \mu_w) = o(t)$ as $t\rightarrow \infty$ in \eqref{cor36_eq1}.

	When we treat the control system in (\ref{Gen_Ctrl_Dyn}) and (\ref{Gen_Ctrl_Ctrl}) as a feedback communication channel and the stabilizing control law as a coding scheme of the feedback channel, as \hyperref[sec24]{Section 2.4} introduces, a control-theoretic interpretation of \hyperref[thm34]{Theorem 3.4} can then be obtained. To synthesize a mean-square stabilizing controller for the system described by (\ref{Gen_Ctrl_Dyn}), (\ref{Gen_Ctrl_Ctrl}) and \hyperref[fig1]{Fig.~1}, the (supremum of) `control' rate $\bar{I}(e; x_0, c)$ (or $\mathcal{C}_f$) achieved by the controller must be large enough to i) offset the instability rate of plant $\bar{I}(e; x_0)$ or $\bar{I}(e; x_0, c)$, and ii) counteract the divergence rate $\bar{D}(\mu_e \| \mu_w)$ induced by the control noise $w$. The first requirement is a fundamental limitation for the controller to stabilize the continuous-time plant model in \eqref{Gen_Ctrl_Dyn}, as we will later show that the instability rate $\bar{I}(e; x_0)$, similar to the Bode-type integrals and data rate, is equal or bounded below by the sum of open-loop unstable poles in the LTI scenario, and analogous to the entropy cost function and LTV Bode's integral, is bounded below by the sum of spectral values of the antistable component in the LTV scenario. The second requirement is a fundamental constraint for the controller to attenuate the external disturbance. When $\bar{D}(\mu_e \| \mu_w) = 0$, all control effort is allocated to stabilize the plant. Thus far, we have interpreted the trade-off properties of total information rate $\bar{I}(e; x_0, c)$ in \hyperref[thm34]{Theorem 3.4} and provided a direct method to compute $\bar{I}(e; x_0, c)$ in \hyperref[thm31]{Theorem 3.1}. A vanilla example in \hyperref[appA]{Appendix A} provides an intuitive preview for the calculation and analysis of total information rate as a metric characterizing Bode's fundamental limitation in an LTI scalar system. More rigorous treatments are given in the following subsections, in which \hyperref[thm31]{Theorem 3.1} is utilized to compute and analyze the information rates, $\bar{I}(e; x_0, c)$ and $\bar{I}(e; x_0)$, in \hyperref[thm34]{Theorem 3.4} with various control systems subject to LTI, LTV, and nonlinear plants. 
	\begin{rem}
		\hyperref[thm31]{Theorems 3.1}, \ref{thm34}, and \hyperref[prop32]{Proposition 3.2} can be applied to all continuous-time control systems depicted in \hyperref[sec24]{Section 2.4} despite the linearity of plant, stationarity and distributions of the input, output and commanded signals, and the distribution of initial state. In the absence of command signal $c_0^t$, total information rate $\bar{I}(e; x_0, c)$ coincides with $\bar{I}(e; x_0)$, and \hyperref[thm31]{Theorem 3.1} and \hyperref[prop32]{Proposition 3.2} remain valid, while the second inequalities in \eqref{prop34_eq1} and \eqref{cor36_eq1} of \hyperref[thm34]{Theorem 3.4} reduce to equalities. 
	\end{rem}

	\subsection{Control Trade-offs in LTI Systems}\label{sec32}
	We now narrow down our scope to the LTI systems descried as follows
	\begin{equation}\label{LTI_Model}
		\begin{split}
			dx_t & = Ax_t dt + B de_t \allowdisplaybreaks  \\
			y_t & = Cx_t, 
		\end{split}
	\end{equation}
	where $A$, $B$ and $C$ are time-invariant matrices of proper dimensions. When the control mapping $\mathcal{K}$ in \hyperref[fig1]{Fig.~1} is linear, such as the state feedback, output feedback, and observer-based feedback controllers, we use \eqref{LTI_Model} to represent the augmented dynamics of the plant $\mathcal{P}$ and control mapping $\mathcal{K}$ such that the control signal can be expressed as
	\begin{equation}\label{Linear_Ctrl}
		u_t = {K}{x}_t,
	\end{equation}
	where $K$ is the control gain, and $x_t$ is the states of the augmented dynamics $\mathcal{U}$ in \eqref{LTI_Model}. By applying \hyperref[thm31]{Theorem~3.1} to the control channel depicted by \eqref{Gen_Ctrl_Err}, \eqref{LTI_Model}, and \eqref{Linear_Ctrl}, an equality between the information rate $\bar{I}(e; {x}_0)$ in \hyperref[thm34]{Theorem~3.4} and the sum of open-loop unstable poles is established for the first time in continuous-time LTI control systems. 
	\begin{prop}\label{prop36}
		When the closed-loop system subject to the augmented LTI plant \eqref{LTI_Model} and linear controller \eqref{Linear_Ctrl} as in \hyperref[fig1]{Fig.~1} is internally mean-square stable, the mutual information rate satisfies \vspace{-0.25em}
		\begin{equation}\label{prop36_eq1}
			\bar{I}(e; {x}_0)  = \sum_{i} \lambda^{+}_i({A}), \vspace{-0.5em}
		\end{equation}  
		where $x_0$ denotes the initial states in \eqref{LTI_Model}, and $\lambda^{+}_i(A)$ are the eigenvalues of matrix $A$ with positive real parts, i.e., open-loop unstable poles of $\mathcal{U}$. 
	\end{prop} \vspace{-1.5em}
	\begin{pf}
		Since the analysis and results of the LTI systems can be treated as a special case of later LTV analysis and results, we postpone and combine the proof of \hyperref[prop36]{Proposition~3.6} with the proof of \hyperref[prop311]{Proposition~3.11}. \hspace*{\fill}$\blacksquare$
	\end{pf} \vspace{-1.25em}
	
	{By virtue of the I-MMSE relationships or \hyperref[thm31]{Theorem 3.1}, an equality between total information (or plant's instability) rate and the sum of open-loop unstable poles is established in \hyperref[prop36]{Proposition~3.6}, which indicates that in addition to the control trade-off properties interpreted in \hyperref[thm34]{Theorem~3.4}, $\bar{I}(e; x_0)$ also serves as an information-theoretic interpretation for the classical Bode's integral of LTI control systems considered by \cite{Bode_1945} and \cite{Freudenberg_TAC_1985}. Previously, in the study of other information-theoretic trade-offs, e.g., Bode-like integral, this equality was missing and impossible to attain by using the mean-square stability or maximum entropy (Lemma~4.1 of \citealp{Li_TAC_2013}). Meanwhile, \hyperref[prop36]{Proposition~3.6} does not postulate any restriction on the stationarity or distribution of signals and initial states. It is worth noting that same as the classical results on Bode's integral \citep{Freudenberg_TAC_1985, Wu_TAC_1992}, the right-hand-side (RHS) sum in \eqref{prop36_eq1} contains the open-loop unstable poles of both the plant $\mathcal{P}$ and control mapping $\mathcal{K}$. When we exclude the unstable poles of control mapping or minimize the RHS sum by choosing a proper control mapping, \eqref{prop36_eq1} then implies that $\bar{I}(e; {x}_0)$ must be lower-bounded by the sum of unstable poles in $\mathcal{P}$ or the sum of unstable poles in $\mathcal{P}$ and $\mathcal{K}$.}
	
	\vspace{-0.25em}

	For completeness, we also provide a lower bound for the total information and instability rates when the LTI plant \eqref{LTI_Model} is stabilized by a causal nonlinear controller \eqref{Gen_Ctrl_Ctrl} in the following lemma, which can be implied from \hyperref[thm34]{Theorem~3.4} and Lemma 4.1 of \citealp{Li_TAC_2013}. Although the exact values of $\bar{I}(e; x_0, c)$ and $\bar{I}(e; x_0)$ can still be computed by using \hyperref[thm31]{Theorem~3.1}, since the control mapping $\mathcal{K}$ is now a nonlinear process, we should resort to the nonlinear estimation method, which will be discussed in \hyperref[sec34]{Section~3.4}. \vspace{-0.5em}
	\begin{lem}\label{lem37}
		When the closed-loop systems subject to the LTI plant~\eqref{LTI_Model} and nonlinear control mapping \eqref{Gen_Ctrl_Ctrl}, as \hyperref[fig1]{Fig.~1} shows, are internally mean-square stable, the information rates $\bar{I}(e; x_0, c)$ and $\bar{I}(e; x_0)$ satisfy
		\begin{equation}\label{prop310_eq1}
			\bar{I}(e; x_0, c)  \geq \bar{I}(e; x_0)  \geq \sum_{i} \lambda^{+}_i(A) ,
		\end{equation}
		where $x_0$ denotes the initial states in \eqref{LTI_Model}, and $\lambda^{+}_i(A)$ are the eigenvalues of matrix $A$ with positive real parts. 
	\end{lem}

	\hyperref[prop36]{Proposition~3.6} and \hyperref[lem37]{Lemma~3.7} respectively develop an equality and an inequality between the total information (or instability) rate and the sum of unstable poles. By combining these results with \eqref{prop34_eq1} and \eqref{cor36_eq1}, an LTI explanation or instance can then be assigned to the control limitation in \hyperref[thm34]{Theorem~3.4}: The total information rate $\bar{I}(e; x_0, c)$ of any causal controller that stabilizes the LTI plant \eqref{LTI_Model} must exceed the sum of the open-loop unstable poles, which suggests that $\bar{I}(e; x_0, c)$ and $\bar{I}(e; x_0)$ serve a similar role as the Bode-type integrals \citep{Bode_1945, Freudenberg_TAC_1985, Li_TAC_2013} in the continuous-time control systems with LTI plants. While this LTI result seems anticipated and shares some similarities with a few other information-theoretic constraints, e.g., SNR constraint \citep{Braslavsky_TAC_2007}, and data rate constraint of discrete-time control systems \citep{Nair_PIEEE_2007}, in the following subsections, we will show that by using \hyperref[thm31]{Theorem~3.1}, the analysis of $\bar{I}(e; x_0, c)$ and $\bar{I}(e; x_0)$ as a control limitation can be easily carried over to the time-varying and nonlinear systems, which are challenging for the traditional methods and previously received less attention.

	\subsection{Control Trade-offs in LTV Systems}
	Consider the LTV plant $\mathcal{P}$ described as follows
	\begin{equation}\label{LTV_Plant}
		\begin{split}
			dx_t & = A(t) x_t dt + B(t) de_t\\
			y_t &= C(t)x_t,
		\end{split}
	\end{equation}
	where $A(t)$, $B(t)$ and $C(t)$ are matrix functions of proper dimensions. A brief introduction on the LTV-related notions, such as the Lyapunov exponents, dichotomy spectrum, sensitivity operator, and inner/outer factorization, is given in \hyperref[AppB]{Appendix B}, and more detailed explanations can be found in the work of \cite{Dieci_JNA_1997} and \cite{Iglesias_LAA_2002}. The following assumptions and constraints are imposed on \eqref{LTV_Plant}. 
	
	\noindent (A5)\label{ass5} The matrix function $A(t): \mathbb{R}^+ \rightarrow \mathbb{R}^{n\times n}$ admits an exponential dichotomy of rank $n_s$, and the antistable component has dichotomy spectrum
	\begin{equation}\label{dich_spec}
		[\underline{\lambda}_1, \overline{\lambda}_1] \cup [\underline{\lambda}_2, \overline{\lambda}_2] \cup \cdots \cup [\underline{\lambda}_{l}, \overline{\lambda}_{l}]  
	\end{equation}
	with dimensions $n_1, \cdots, n_l$, and $\sum_{i=1}^{l} n_i = n_u$, where $n_s + n_u = n$.
	
	\noindent (A6)\label{ass6} $A(t) - B(t)C(t)$ is uniformly exponentially stable (UES). Equivalently, the sensitivity operator $S = S_i S_o$ of \eqref{LTV_Plant}, the inner factor $S_i$, the outer factor $S_o$ and its inverse $S_o^{-1}$, are all bounded. 
	
	\noindent (A7)\label{ass7} The relative degree of the open-loop system $\mathcal{P}$ is at least $2$, i.e., $C(t)B(t) = 0$ for all $t$. 
	
	\begin{rem}
		Similar to the stable/unstable dichotomy of LTI systems, (\hyperref[ass5]{A5}) is a fundamental property of LTV systems that has been extensively studied and adopted in the existing literature \citep{Coppel_1978, Dieci_JDE_2010, Tranninger_CSL_2020}. When \eqref{LTV_Plant} degenerates to an LTI system, the intervals in \eqref{dich_spec} shrink to the points that coincide with the real parts of the LTI system's unstable poles. (\hyperref[ass6]{A6}) is the prerequisite of any internally stabilizing controller~\citep{Iglesias_LAA_2002}. Previous investigations of Bode's integral in LTI systems require the open-loop systems have a relative degree of at least 2 or $CB = 0$ \citep{Seron_1997}, and (\hyperref[ass7]{A7}) is the counterpart constraint in LTV systems. 
	\end{rem}

	\noindent When $A(t)$ admits an exponential dichotomy, analogous to the modal decomposition on LTI models, we have the following stability preserving state-space transformation that separates the stable and antistable parts of $A(t)$ \citep[Chapter 5]{Coppel_1978}.
	
	\begin{lem}\label{lem39}
		When the matrix function $A(t)$ in \eqref{LTV_Plant} admits an exponential dichotomy, there exists a bounded matrix function $T(t)$ with bounded inverse such that  
		\begin{equation}\label{lem220_eq1}
			\left[\begin{array}{c;{3pt/2pt}c}
				\varTheta(t) & T(t)B(t) \\ \hdashline[3pt/2pt]
				C(t)T^{-1}(t) & D(t)
			\end{array}  \right] 
			:=
			\left[\begin{array}{cc;{3pt/2pt}c}
				A_s(t) & 0 & B_s(t)\\
				0 & A_u(t) & B_u(t)\\ \hdashline[3pt/2pt]
				C_s(t) & C_u(t) & D(t)
			\end{array}\right],
		\end{equation}
		where $D(t)$ is the feedthrough matrix; $\varTheta(t) =[\dot{T}(t) + T(t)A(t)] T^{-1}(t)$; $A_s(t)$ is UES, and $A_u(t)$ is uniformly exponentially antistable (UEA).
	\end{lem}
	\noindent By reproducing the relationship between Bode's integral and the LTI entropy cost function, a time-domain integral $\mathcal{B}$ is proposed as an LTV analogue of Bode's integral \citep{Iglesias_LAA_2002}.

	\begin{lem}\label{lem310}
		For the LTV plant \eqref{LTV_Plant} satisfying (\hyperref[ass5]{A5})-(\hyperref[ass7]{A7}), the following integral $\mathcal{B}$ defines a control trade-off 
		\begin{equation*}
			\mathcal{B} := \lim_{t\rightarrow\infty} \frac{1}{4t} \int_{-t}^{t} {\rm{tr}}[\hat{S}_o(\tau, \tau_0)] d\tau   = \lim_{t\rightarrow \infty} \frac{1}{2t} \int_{-t}^{t} {\rm{tr}}[A_u(\tau)] d\tau, 
		\end{equation*} 
		where $\hat{S}_o(\tau, \tau_0) = [B^\top(\tau) X(\tau) - C(\tau)]\varPhi_{A-BC}(\tau, \tau_0) B(\tau_0)$ is the kernel of the outer factor $S_o$ with $\tau \geq \tau_0$,  $X(\tau) = X^\top(\tau) \geq 0$ being a stabilizing solution to the Riccati differential equation $-\dot{X}(\tau) = A^\top(\tau)X(\tau) + X(\tau) A(\tau) - X(\tau)B(\tau)B^\top(\tau)X(\tau)$, and $\varPhi_{A-BC}(\tau, \tau_0)$ standing for the state transition matrix of $A(t) - B(t)C(t)$ from $\tau_0$ to $\tau$. Meanwhile, the control trade-off integral $\mathcal{B}$ satisfies
		\begin{equation*}
			0 < \sum_{i=1}^{l} n_i \underline{\lambda}_i \leq \lim_{t\rightarrow \infty} \frac{1}{2t} \int_{-t}^{t}  {\rm{tr}}[A_u(\tau)] d\tau \leq \sum_{i=1}^{l} n_i \overline{\lambda}_i,
		\end{equation*}
		where $A_u(\tau)$ is the antistable part of $A(\tau)$, and $\underline{\lambda}_i$ and $\overline{\lambda}_i$, $i =1, \cdots, l$, are the spectral values of $A_u(\tau)$.
	\end{lem}

	When the stabilizing controller $\mathcal{K}$ is linear, we use \eqref{LTV_Plant} to describe the augmented or cascade dynamics of the plant $\mathcal{P}$ and control mapping $\mathcal{K}$ in \hyperref[fig1]{Fig. 1(b)} such that the control signal satisfies
	\begin{equation}\label{LTV_LinearCtrl}
		u(t) = K(t)x(t),
	\end{equation}
	where $K(t)$ is the time-varying control gain. By applying \hyperref[thm31]{Theorem 3.1} to the control system or channel governed by \eqref{Gen_Ctrl_Err}, \eqref{LTV_Plant} and \eqref{LTV_LinearCtrl}, we obtain the following relationship for $\bar{I}(e; x_0)$ in the LTV scenario. 
	\begin{prop}\label{prop311}
			When the closed-loop system subject to the augmented LTV plant \eqref{LTV_Plant} and linear control mapping \eqref{LTV_LinearCtrl}, as in \hyperref[fig1]{Fig. 1}, is internally mean-square stable, the mutual information rate $\bar{I}(e; x_0)$ satisfies
			\begin{align} \label{prop312_eq1}
				\hspace{-10pt} \bar{I}(e; x_0) & =  \lim_{t \rightarrow \infty} \frac{1}{t} \int_{0}^{t} {\rm tr}[A_u(\tau)] d\tau \allowdisplaybreaks \\
				& \hspace{48pt} - \lim_{t \rightarrow \infty} \frac{1}{2t}  \int_{0}^{t} {\rm tr}[\dot{{P}}_u(\tau){P}_u^{-1}(\tau)] d\tau \nonumber \allowdisplaybreaks \\
				& \geq \sum_{i=1}^{l} n_i \underline{\lambda}_i  - \limsup_{t \rightarrow \infty} \frac{1}{2} {\rm tr}[\dot{{P}}_u(t){P}_u^{-1}(t)], \nonumber
			\end{align}
			\noindent where $A_u(\tau)$ is the antistable part of $A(\tau)$; $P_u(\tau) = \mathbb{E}[(x_u(\tau) - \hat{x}_u(\tau))(x_u(\tau) - \hat{x}_u(\tau))^\top ]$ is the estimation covariance of the antistable modes, and $\{\underline{\lambda}_i\}_{i=1}^l$, with dimensions $n_1, \cdots, n_l$, are the spectral values of $A_u(\tau)$.
	\end{prop} \vspace{-1em}
	\begin{pf}
		To restore the proof for \hyperref[prop36]{Proposition 3.6}, we only need to replace the following time-varying terms from \eqref{LTV_Plant} and \eqref{LTV_LinearCtrl} by the time-invariant terms in \eqref{LTI_Model} and \eqref{Linear_Ctrl}. 
		By applying \hyperref[thm31]{Theorem 3.1} to the LTV systems described by \eqref{LTV_Plant} and \eqref{LTV_LinearCtrl}, we have
		\begin{equation*}\label{prop317_eq2}
			\bar{I}(e; x_0) = \lim_{t \rightarrow \infty} \frac{1}{2t} \mathbb{E}[\|u - \hat{u}\|^2],
		\end{equation*}
		where $u_\tau = K_\tau x_\tau$ and $\hat{u}_\tau = \mathbb{E}[u_\tau | e_0^\tau]$. Hence, to prove the equality in \eqref{prop312_eq1}, we only need to show that $(2t)^{-1} \cdot \mathbb{E}[\|u - \hat{u}\|^2] = t^{-1} \int_{0}^{t} {\rm tr}[A_u(\tau)] d\tau - (2t)^{-1}  \int_{0}^{t} {\rm tr}[\dot{P}_u(\tau) \cdot P_u^{-1}(\tau)]d\tau$. When computing the estimate $\hat{u}_\tau$, we resort to the following LTV filtering problem obtained by substituting control mapping \eqref{LTV_LinearCtrl} into LTV plant \eqref{LTV_Plant} and control channel \eqref{Gen_Ctrl_Err}, respectively: 
		\begin{equation}\label{prop317_eq3}
			\begin{split}
				dx_t & = \bar{A}(t)x_t dt + B(t)dw_t\\
				de_t & = K(t) x_t dt + dw_t,
			\end{split}
		\end{equation}
		where $\bar{A}(t) = A(t) + B(t) K(t)$. Let the estimation error covariance matrix be $P_t = \mathbb{E}[(x_t - \hat{x}_t)(x_t - \hat{x}_t)^\top]$, where $\hat{x}_t = \mathbb{E}[x_t|e_0^t]$. We can then solve $P_t$ from the following Riccati differential equation or RDE \citep[Theorem~10.3]{Liptser_2001}
		\begin{equation*}
			\dot{P}^{}_t = \bar{A}^{}_t P^{}_t + P^{}_t \bar{A}_t^\top + B^{}_t B_t^\top - (B^{}_t + P^{}_tK_t^\top)(B^{}_t + P^{}_tK_t^\top)^\top,
		\end{equation*}
		or equivalently 
		\begin{equation}\label{prop317_eq4}
			\dot{P}_t = A_t P_t + P_t A_t^\top  - P_tK_t^\top K_t P_t,
		\end{equation}
		which is also the RDE of \eqref{prop317_eq3} without the control input:
		\begin{equation}\label{prop317_eq5}
			\begin{split}
				dx_t & = A(t) x_t dt \allowdisplaybreaks\\
				de_t & = K(t) x_t dt + dw_t. \allowdisplaybreaks
			\end{split}
		\end{equation}
		When the LTV system \eqref{prop317_eq5} is exponentially stable or uniformly completely reconstructible \citep{Sivan_1972, Ni_SCL_2016}, given any initial condition $P(t_0) \geq 0$, the solution of \eqref{prop317_eq4} converges to $P(t)$, i.e., $P(t) = \lim_{\tau \rightarrow \infty} P(\tau)$. By invoking \hyperref[lem39]{Lemma 3.9} (or the stable/unstable decomposition for LTI case), we can transform \eqref{prop317_eq5} into the following modal form
		\begin{equation}\label{prop317_eq6}
			\begin{split}
				\left[\begin{matrix}
					dx_s(t)\\
					dx_u(t)
				\end{matrix}\right] & = \left[\begin{matrix}
					A_s(t) & 0\\
					0 & A_u(t)
				\end{matrix}\right] \left[  \begin{matrix}
					x_s(t)\\
					x_u(t)
				\end{matrix}\right]dt \\
				de_t & = \left[ \begin{matrix}
					K_s(t) & K_u(t)
				\end{matrix}\right] \left[\begin{matrix}
					x_s(t)\\
					x_u(t)
				\end{matrix}\right]dt + dw_t,
			\end{split}
		\end{equation}
		where $A_s(t)$ is UES, and $A_u(t)$ is UEA. Let $\bar{K}(t) = K(t)T^{-1}(t) = [K_s(t)   \ K_u(t)]$ and $\bar{x}
		(t) = [x_s^\top(t) \ x_u^\top(t)]^\top$. Based on \hyperref[lem46]{Lemma 4.6} in \hyperref[sec4]{Section 4}, when the LTV system \eqref{prop317_eq5} and thus also \eqref{prop317_eq6} are uniformly completely reconstructible, the estimation error covariance matrix $\bar{P}(t) = \mathbb{E}[(\bar{x}_t - \hat{\bar{x}}_t)(\bar{x}_t - \hat{\bar{x}}_t)^\top]$ of \eqref{prop317_eq6} takes the form of $\bar{P}(t) = \textrm{diag}\{0, P_u(t)\}$, where $P_u(t)$ is the positive definite solution of the following RDE:
		\begin{align}\label{prop317_eq7}
			\dot{P}_u(t) = A_u(t) P_u(t) & + P_u(t) A_u^\top(t)  \\
			& - P_u(t)K^\top_u(t)K_u(t)P_u(t). \nonumber
		\end{align}
		Meanwhile, since $u_t = \bar{K}_t\bar{x}_t$, ${\rm tr}(\bar{P}_t) = {\rm tr}[P_u(t)]$, and ${\rm tr}(\bar{K}^{}_t\bar{P}^{}_t\bar{K}^\top_t) = {\rm tr}[K_u^{}(t) P^{}_u(t) K^\top_u(t)]$, the MS estimation errors satisfy
		\begin{subequations}
			\begin{align}
				\textrm{tr}[{P}_u(t)]  = \mathbb{E}[ (\bar{x}_t - \hat{\bar{x}}_t)^\top(\bar{x}_t - \hat{\bar{x}}_t)  ], \allowdisplaybreaks\\
				\textrm{tr}[{K}^{}_u(t) {P}^{}_u(t) {K}^\top_u(t)] = \mathbb{E}[(u_t - \hat{u}_t)^\top(u_t - \hat{u}_t)].\label{prop312_eq8b}
			\end{align}
		\end{subequations}
		Integrating both sides of \eqref{prop312_eq8b} from $\tau = 0$ to $t$ gives
		\begin{align}\label{prop312_eq9}
			\int_{0}^{t}\textrm{tr}[{K}^{}_u(\tau){P}_u(\tau){K}^\top_u(\tau)]d\tau  = \mathbb{E}[\|u - \hat{u}\|^2] , 
		\end{align}
		where Fubini's theorem is used to exchange the order of integration and expectation. Pre- and post-multiplying \eqref{prop317_eq7} by $P_u^{-1/2}(\tau)$ and substituting it into \eqref{prop312_eq9} yields
		\begin{align}\label{prop317_eq8}
			\mathbb{E}[\|u - \hat{u}\|^2] & = \int_{0}^{t} {\rm tr}[  P^{1/2}_u(\tau)K^\top_u(\tau) K^{}_u(\tau) P^{1/2}_u(\tau) ]d\tau \nonumber \nonumber \allowdisplaybreaks \\
			& =   \int_{0}^{t}{\rm tr}[2 A_u(\tau) - \dot{P}_u(\tau)P^{-1}_u(\tau) ] d\tau, 
		\end{align}
		where the second equality follows from $P^{1/2}_u(\tau)K^\top_u(\tau) \cdot K^{}_u(\tau)  P^{1/2}_u(\tau)  = P_u^{-1/2}(\tau)   A_u(\tau)P_u^{1/2}(\tau) + P_u^{1/2}(\tau)A_u^\top(\tau) \cdot P_u^{-1/2}(\tau) - P_u^{-1/2}(\tau) \dot{P}_u(\tau)P_u^{-1/2}(\tau)$ obtained from the preceding manipulation on \eqref{prop317_eq7}. From \eqref{prop317_eq8}, we can then derive \eqref{prop312_eq1} by
		\begin{align*}
			\bar{I}(e; x_0) & = \lim_{t \rightarrow \infty} \frac{1}{2t}   \mathbb{E}[\|u - \hat{u}\|^2] \allowdisplaybreaks \\
			& = \lim_{t \rightarrow \infty} \frac{1}{2t} \int_{0}^{t} {\rm tr}[ 2 A_u(\tau) - \dot{P}_u(\tau)P^{-1}_u(\tau)] d\tau  \allowdisplaybreaks \\
			& \geq \sum_{i=1}^{l} n_i \underline{\lambda}_i  - \limsup_{t \rightarrow \infty} \frac{1}{2} {\rm tr}[\dot{{P}}_u(t){P}_u^{-1}(t)], \allowdisplaybreaks
		\end{align*}
		where the inequality follows from \hyperref[lem310]{Lemma~3.10}, the convergence of $P_u(t) = \lim_{\tau \rightarrow \infty}P_u(\tau)$, and the application of L'H\^opital's rule, $\lim_{t\rightarrow\infty} \int_{0}^{t} \textrm{tr}[\dot{P}^{}_u(\tau)P_u^{-1}(\tau)]  d\tau / t = \lim_{t \rightarrow \infty} {\rm tr}[\dot{P}^{}_u(t)P_u^{-1}(t)] \leq \limsup_{t \rightarrow \infty} {\rm tr}[\dot{P}^{}_u(t)P_u^{-1}(t)]$. Moreover, by comparing \eqref{prop312_eq1} and \eqref{prop316_eq1} below, we can also conclude that $\limsup_{t \rightarrow \infty}  {\rm tr}[\dot{{P}}_u(t){P}_u^{-1}(t)] \leq 0$ in \eqref{prop312_eq1}. This completes the proof. \hspace*{\fill}$\blacksquare$
	\end{pf} \vspace{-1em}

	For completeness, we also consider the scenario when the LTV plant \eqref{LTV_Plant} is stabilized by a causal nonlinear controller \eqref{Gen_Ctrl_Ctrl}. Unlike \hyperref[prop311]{Proposition 3.11}, wherein an equality of $\bar{I}(e; x_0)$ is attained by using \hyperref[thm31]{Theorem 3.1} or Duncan's theorem, the following lemma, based on the mean-square stability and maximum entropy, only provides a lower bound for $\bar{I}(e; x_0, c)$ and $\bar{I}(e; x_0)$.
	
	\vspace{-0.5em}
	
	\begin{cor}\label{cor312}
		When the closed-loop system, subject to the LTV plant \eqref{LTV_Plant} and nonlinear controller \eqref{Gen_Ctrl_Ctrl} as in \hyperref[fig1]{Fig.~1}, are internally mean-square stable, we have
		\begin{align}\label{prop316_eq1}
			\bar{I}(e; x_0, c) \geq \bar{I}(e; x_0) & \geq \lim_{t\rightarrow \infty} \frac{1}{t} \int_{0}^{t}{\rm tr}[A_u(\tau)]d\tau \nonumber \allowdisplaybreaks \\
			&  \geq \sum_{i=1}^{l} n_i \underline{\lambda}_i ,
		\end{align}
		where $A_u(\tau)$ is the antistable part of $A(\tau)$, and $\{\underline{\lambda}_i\}_{i=1}^{l}$, with dimensions $n_1, \cdots, n_l$, are the spectral values of the antistable component $A_u(\tau)$.
	\end{cor}	\vspace{-1.5em}
	\begin{pf}
		See \hyperref[AppC]{Appendix C} for the proof.  \hspace*{\fill}$\blacksquare$
	\end{pf}\vspace{-1.5em}
	\noindent Since the lumped input process $\mathcal{U}$ in \eqref{Gen_Ctrl_Err} and \hyperref[fig1]{Fig. 1} is no longer linear in the setup of \hyperref[cor312]{Corollary 3.12}, for the exact values of $\bar{I}(e; x_0, c)$ and $\bar{I}(e; x_0)$, we should resort to \hyperref[thm31]{Theorem 3.1} and the nonlinear estimation method to be discussed in the next subsection.

	\hyperref[prop311]{Proposition 3.11} and \hyperref[cor312]{Corollary 3.12} for the first time present an equality and an inequality between the total information (or instability) rate and the sum of spectral values of the antistable component or the LTV Bode's integral $\mathcal{B}$ in the continuous-time control systems described by \eqref{LTV_Plant}. It is worth noting that similar to the RHS sum in \eqref{prop36_eq1} and the lower bound of integral $\mathcal{B}$ in \hyperref[lem310]{Lemma 3.10}, the lower bound in \eqref{prop312_eq1} is determined by the unstable portions of both plant and controller dynamics. Substituting \eqref{prop312_eq1} and \eqref{prop316_eq1} into \eqref{prop34_eq1} and \eqref{cor36_eq1}, we can readily imply that the total information rate attained by any causal controller that stabilizes the LTV plant \eqref{LTV_Plant} must exceed the sum of spectral values. This implication offers an LTV explanation to \hyperref[thm34]{Theorem 3.4} and is consistent with the preceding interpretations for the LTI and general control systems. More specifically, owing to the relationship between $\bar{I}(e; x_0)$ and $\mathcal{B}$ in \eqref{prop312_eq1} and \eqref{prop316_eq1}, the total information (or instability) rate serves as a time-domain trade-off, similar to the LTV Bode's integral \citep{Iglesias_LAA_2002} and the entropy cost function of minimum entropy control problem \citep{Glover_SCL_1988}, in the continuous-time control systems subject to the LTV plant \eqref{LTV_Plant}. When the running control cost is smaller than $\bar{I}(e; x_0, c)$ and $\bar{I}(e; x_0)$ on a time interval, the running cost outside this interval must be larger, and vice versa, which is also analogous to the trade-off property of $\bar{I}(e; x_0, c)$ and Bode's integral in the frequency range of LTI systems.
	
	\vspace{-0.5em}
	
	\begin{rem}\label{rem313}
		\hyperref[prop311]{Proposition 3.11} and \hyperref[cor312]{Corollary 3.12} cover \hyperref[prop36]{Proposition 3.6} and \hyperref[lem37]{Lemma 3.7} as a special case. When we apply the former LTV results to the LTI systems described by \eqref{LTI_Model}, the spectral values in \eqref{prop312_eq1} and \eqref{prop316_eq1} will degenerate to the real parts of the open-loop unstable poles in \eqref{prop36_eq1} and \eqref{prop310_eq1}, and ${\rm tr}[\dot{{P}}^{}_u(t){P}_u^{-1}(t)]$ in \eqref{prop312_eq1} will vanish, since the error covariance matrix $P_u(t)$ converges to a constant matrix $P_u$, i.e., $\lim_{t\rightarrow \infty} \dot{P}_u(t) = 0$.
	\end{rem}

	\subsection{Control Trade-offs in Nonlinear Systems}\label{sec34}	
	At the end of this section, we consider the limitation of nonlinear control systems. Substituting the memory-less controller \eqref{Gen_Ctrl_Ctrl} into the nonlinear plant \eqref{Gen_Ctrl_Dyn} and channel \eqref{Gen_Ctrl_Err}, we obtain the following nonlinear filtering problem
	\begin{equation}\label{NL_Ctrl}
		\begin{split}
			dx_t & = \bar{f}(t, x_t, e_t)dt + b(t, x_t)dw_t \allowdisplaybreaks \\
			de_t & = u(t, e_t)dt + dw_t,
		\end{split}
	\end{equation}
	where $x_t \in \mathbb{R}^n$ denotes the hidden states; the error signal $e_t$ is observable, and $\bar{f}(t, x_t, e_t) = f(t, x_t) + b(t, x_t)  u(t, e_t)$. For conciseness, we first consider the single-input and single-output (SISO) case, i.e., $u_t$ and $e_t\in\mathbb{R}$ in \eqref{NL_Ctrl}, and omit the observable commanded signal $c_t$ in \eqref{Gen_Ctrl_Ctrl}. 
	
	\vspace{-0.5em}
	
	To compute the information rate $\bar{I}(e; x_0)$ from \eqref{NL_Ctrl}, we use the equality $I(e; x_0) = ({1}/{2}) \cdot (\mathbb{E}[\|u\|^2] - \mathbb{E}[\|\hat{u}\|^2])$ from \hyperref[thm31]{Theorem 3.1}, where the first expectation $\mathbb{E}[\|u\|^2]$ is usually observable, and the second expectation $\mathbb{E}[\|\hat{u}\|^2]$ and $\hat{u}(\tau, e_0^\tau, x_0) = \mathbb{E}[u(\tau, e_0^\tau, x_0)|e_0^\tau]$ need to be estimated from \eqref{NL_Ctrl}. When $u(t, x_t)$ is not observable, we can follow the scheme in \hyperref[sec44]{Section 4.4} to calculate $\bar{I}(e; x_0)$. To implement a nonlinear filter on \eqref{NL_Ctrl}, analogous to the stabilizability and detectability conditions for solving ARE and RDE, additional regularity and differentiability constraints are required to guarantee the existence and uniqueness of solution to \eqref{NL_Ctrl} and the pathwise uniqueness of filter equation \citep{Mitter_1982, Liptser_2001, Lucic_AAP_2001, Xiong_2008, Bain_2009}. Let $k(t, x, e)$ denote any of the functions $\bar{f}(t, x_t, e_t)$, $b(t, x_t)$ and $u(t, e_t)$ in \eqref{NL_Ctrl}. We require $k(t, x, e)$ be Lipschitz, i.e., $|k(t, x_1, e_1) - k(t, x_2, e_2)|^2 \leq K_1(|x_1 - x_2|^2 + |e_1 - e_2|^2)$ for some constant $K_1 > 0$, and $k^2(t, x, e) \leq K_2(1+x^2 + e^2)$ for some constant $K_2 > 0$. Meanwhile, the control input $u(t, e_t)$ needs to be at least once differentiable at $t$ and twice differentiable at $e_t$ such that $\sup_{\tau \leq t} \mathbb{E}[u^2(\tau, e_\tau)] < \infty$,  $\int_{0}^{t}\mathbb{E}\left[ (u'_{e_\tau}(\tau, e_\tau))^2 \right]d\tau < \infty$, and $\int_{0}^{t}\mathbb{E}[\mathcal{L}u(\tau, e_\tau)]^2 d\tau < \infty$, where $\mathcal{L}u(\tau, e_\tau) = u'_\tau(\tau,  e_\tau)  +  u'_{e_\tau}(\tau, e_\tau)u(\tau, e_\tau) + u''_{e_\tau e_\tau}(\tau, e_\tau) / 2$. When \eqref{NL_Ctrl} fulfills the preceding regularity and differentiability constraints, based on \hyperref[thm31]{Theorem 3.1} and the Stratonovich-Kushner equation, the following proposition provides an approach to compute the information rate $\bar{I}(e; x_0)$ in nonlinear control systems. \vspace{-0.5em}
	{\begin{prop}\label{prop314}
			When the closed-loop systems described by the nonlinear plant and controller \eqref{NL_Ctrl}, as in \hyperref[fig1]{Fig. 1}, are internally mean-square stable, we have\vspace{-1em}
			\begin{align}\label{prop321_eq1}
				\bar{I}(e; x_0) = & \lim_{t\rightarrow\infty}\frac{1}{2t}   \int_{0}^{t}   \mathbb{E}[u^2(\tau, e_0^\tau, x_0)] d\tau 
				\\
				& \hspace{30pt} - \lim_{t \rightarrow \infty} \frac{1}{2t} \int_{0}^{t} \mathbb{E}[\hat{u}^2(\tau, e_0^\tau, x_0)  ]   d\tau, \nonumber
			\end{align}	
			where $\hat{u}(\tau, e_0^\tau, x_0)  =  \mathbb{E}[u(\tau, e_0^\tau, x_0)|e_0^\tau] = \pi_\tau(u)$ satisfies  \vspace{-0.5em}
			\begin{align}\label{lem39_eq1}
				\pi_\tau(u) = \pi_0(u) & + \int_{0}^{\tau}\pi_s(\mathcal{L}u)ds \allowdisplaybreaks \\
				& + \int_{0}^{\tau} \left[ \pi_s(u'_{e_s}) + \pi_s(u^2) - \pi^2_s(u) \right] d\tilde{w}_s ; \nonumber
			\end{align}
			$\mathcal{L}u(s, e_s) = u'_s(s, e_s) +  u'_{e_s}(s, e_s)u(s, e_s) + \frac{1}{2} u''_{e_s e_s}(s, e_s)$, and $\tilde{w}_\tau = \int_{0}^{\tau}de_s - \int_{0}^{\tau} \pi_s(u)ds$ is a Wiener process w.r.t. $\mathcal{F}_\tau^e$, $\tau \in [0, t]$.
	\end{prop}}\vspace{-1em}
	\begin{pf}
		Applying \hyperref[thm31]{Theorem 3.1} to the control channel in \eqref{NL_Ctrl}, we have
		\begin{align}
			\bar{I}(e; x_0) & \neweq{(a)} \lim_{t \rightarrow \infty} \frac{1}{2t} \left( \mathbb{E}[\|u\|^2] - \mathbb{E}[\|\hat{u}\|^2]  \right) \nonumber \allowdisplaybreaks\\
			& \neweq{(b)} \lim_{t \rightarrow \infty} \frac{1}{2t}  \mathbb{E}\left[ \int_{0}^{t} u^2(\tau, e_0^\tau, x_0) d\tau \right] \nonumber \allowdisplaybreaks\\
			& \hspace{50pt} -  \lim_{t \rightarrow \infty} \frac{1}{2t}  \mathbb{E}\left[ \int_{0}^{t} \hat{u}^2(\tau, e_0^\tau, x_0) d\tau \right] , \nonumber \allowdisplaybreaks
		\end{align}
		where (a) is obtained by applying \eqref{thm31_eq1} to \eqref{NL_Ctrl}, dividing the result by $t$, and taking the limit as $t\rightarrow \infty$; (b) expands the norms inside expectations, and \eqref{prop321_eq1} is then obtained by using the Fubini's theorem to swap the order of expectation and integration in (b). While the control signal ${u}(\tau, e_0^\tau, x_0)$ in \eqref{prop321_eq1} is observable, the optimal estimate $\pi_\tau(u) = \hat{u}(\tau, e_0^\tau, x_0) = \mathbb{E}[u(\tau, e_0^\tau, x_0)|e_0^\tau]$ needs to be estimated from the nonlinear filtering problem \eqref{NL_Ctrl}. By applying the one-dimensional Stratonovich-Kushner equation (\citealp{Liptser_2001}, Theorem~8.3) to \eqref{NL_Ctrl}, we obtain the estimate $\hat{u}(\tau, e_0^\tau, x_0)$ in \eqref{lem39_eq1}. \hspace*{\fill}$\blacksquare$
	\end{pf} \vspace{-1.5em}

	Due to the various and complex structures of nonlinear plants and systems, the trade-off properties of $\bar{I}(e; x_0)$ in \eqref{prop321_eq1} as well as its connections to some established control trade-offs, such as the Bode-type integrals and entropy cost function, are not as explicit as the preceding discussions for linear systems. Meanwhile, similar to \hyperref[prop36]{Propositions 3.6} and \ref{prop311}, the value of $\bar{I}(e; x_0)$ in \hyperref[prop314]{Proposition 3.14} is determined by both the nonlinear plant and controller. However, by combining \hyperref[prop314]{Proposition 3.14} with \hyperref[thm34]{Theorem 3.4} or \hyperref[prop32]{Proposition 3.2}, we can still glimpse some trade-off features of $\bar{I}(e; x_0)$ in nonlinear control systems. Consistent with the interpretations for linear systems, a straightforward observation from \hyperref[prop314]{Proposition 3.14} is that the total information rate of any causal controller that stabilizes the nonlinear plant \eqref{NL_Ctrl} must exceed the value of $\bar{I}(e; x_0)$ given in \eqref{prop321_eq1}, which can be used as a criterion to examine whether a causal controller is able to stabilize the nonlinear system \eqref{NL_Ctrl}. Meanwhile, since $D(\mu_{e|(x_0, c)} \| \mu_w) = \mathbb{E}[\|u\|^2 / 2]$ and $D(\mu_{e|(x_0, c)} \| \mu_w) = I(e_0^t; x_0, c_0^t) + D(\mu_e\|\mu_w)$ by \hyperref[prop32]{Proposition 3.2}, the total information rate $\bar{I}(e; x_0, c)$ or $\bar{I}(e; x_0)$ also serves as a time-domain trade-off, similar to the entropy cost function and LTV Bode's integral $\mathcal{B}$, in the nonlinear setting. When the quadratic control cost $\mathbb{E}[u_\tau^\top u_\tau]$ is smaller than the value of $\bar{I}(e; x_0)$ in \eqref{prop321_eq1} or, more precisely, the value of $\bar{I}(e; x_0, c) + \bar{D}(\mu_e \| \mu_w)$ on a time interval, the control cost outside this interval must be larger, and vice versa. Up until now, we have analyzed and computed total information rate as a fundamental control trade-off in the continuous-time control systems subject to LTI, LTV, and nonlinear plants. In the following section, total information rate as a fundamental filtering limit will be investigated. 
	
	\vspace{-0.5em}
	
	\begin{rem}\label{rem315}
		To compute the mutual information rate $\bar{I}(e; x_0)$ of a multi-input and multi-output (MIMO) control channel, we only need to rewrite \eqref{NL_Ctrl} and \eqref{lem39_eq1} into a matrix-vector form and replace the one-dimensional filtering equation in \hyperref[prop314]{Proposition 3.14} by the multi-dimensional Stratonovich-Kushner equation (\citealp{Liptser_2001}, Theorem 9.3).
	\end{rem}

	\vspace{-0.25em}
	
	\section{Fundamental Limitations of Filtering}\label{sec4}
	
	\vspace{-0.25em}

	For the filtering problem, we first show that total information rate $\bar{I}(y; x_0, w)$ or directed information rate $\bar{I}(z\rightarrow y)$ serves as a filtering limit in the general filtering systems depicted by \eqref{Filter_Model} and \hyperref[fig2]{Fig. 2}, and then apply it to capture the fundamental limitations of the filtering systems subject to LTI, LTV, and nonlinear plants, respectively. 
	
	\vspace{-0.5em}
	
	\subsection{General Filtering Limits}\label{sec41} 
	
	\vspace{-0.5em}
	
	Based on Duncan's theorem, the following theorem provides a direct method to calculate the total information (rate) $I(y_0^t; x_0, w_0^t)$ from the MS estimation error of noise-free output $z(t, w_0^t, x_0)$ in \eqref{Filter_Model}.
	{\begin{thm}\label{thm41}
			For the filtering systems described by \eqref{Filter_Model} and \hyperref[fig2]{Fig. 2}, the total information $I(y_0^t; x_0, w_0^t)$ satisfies
			\begin{equation}\label{thm41_eq1}
				I(y_0^t; x_0, w_0^t) =  \frac{ \mathbb{E}\left[ \| z - \hat{z} \|^2\right]}{2}    = \frac{\mathbb{E} [ \| z\|^2 ] - \mathbb{E} [\|\hat{z} \|^2]}{2} ,
			\end{equation}
			where $\hat{z}(\tau, w_0^\tau, x_0) = \mathbb{E}[z(\tau, w_0^{\tau}, x_0)|y_0^\tau]$, and the total information rate is given by
			\begin{equation}\label{thm41_eq2}
				\bar{I}(y; x_0, w) = \lim_{t\rightarrow \infty} \frac{ I(y_0^t; x_0, w_0^t) }{t} .
			\end{equation}
	\end{thm}} \vspace{-2.5em}
	\begin{pf} 
		Since the observation model in \eqref{Filter_Model} is an additive white Gaussian channel without feedback, by applying the non-feedback version of \hyperref[lem26]{Lemma 2.6} to \eqref{Filter_Model}, we obtain \eqref{thm41_eq1}. The total information rate \eqref{thm41_eq2} follows from \hyperref[Def23]{Definition 2.3}.  \hspace*{\fill}$\blacksquare$
	\end{pf} 
	
	\vspace{-1.5em}
		
	\noindent \hyperref[thm41]{Theorem 4.1} shows that regardless of how the filter $\mathcal{F}$ is designed, ${I}(y_0^t; x_0, w_0^t)$ equals half of the time integral of MS estimation error, and $\bar{I}(y; x_0, w)$ equals half of the time-averaged or steady-state MS estimation error. By using the chain rule of mutual information and \hyperref[lem27]{Lemma 2.7}, more insights on $I(y_0^t; x_0, w_0^t)$ can be gained. 
	
	\vspace{-0.5em}
	
	\begin{prop}\label{prop42}
		For the filtering systems described by \eqref{Filter_Model} and \hyperref[fig2]{Fig. 2}, the total information $I(y_0^t; x_0, c_0^t)$ satisfies
		\begin{equation}\label{prop42_eq1}
			I(y_0^t; x_0, w_0^t) = I(y_0^t; x_0) + I(y_0^t; w_0^t| x_0),
		\end{equation}
		and can be expressed as a difference between divergences
		\begin{equation}\label{prop42_eq2}
			I(y_0^t; x_0, w_0^t) = D(\mu_{y|(x_0, w)} \| \mu_v) - D(\mu_y \| \mu_v),
		\end{equation}
		where $\mu_v$ is the Wiener measure induced by the measurement noise $v_0^t$; $D(\mu_{y|(x_0, w)} \| \mu_v) = \mathbb{E}[\|z\|^2 / 2]$, and $D(\mu_y \| \mu_v) = \mathbb{E}[\|\hat{z}\|^2 / 2]$.
	\end{prop}
	
	\vspace{-1.5em}
	
	\begin{pf}
		By applying the chain rule of mutual information to $I(y_0^t; x_0, w_0^t)$, we can prove \eqref{prop42_eq1}. Equation \eqref{prop42_eq2} can be verified by replacing the mutual information and divergences with their definition forms in \hyperref[def21]{Definition 2.1} and $I(x_0^t; y_0^t) := \mathbb{E}_{(x,y)}[\log(d\mu_{xy} / d(\mu_x \times \mu_y))] = \mathbb{E}_y[\log(d\mu_{y|x} /  d\mu_y)]$ in \hyperref[Def23]{Definition 2.3}. Meanwhile, by applying the non-feedback version of \hyperref[lem27]{Lemma 2.7} to the measurement channel \eqref{Filter_Model}, we have $D(\mu_{y|(x_0, w)} \| \mu_v) = \mathbb{E}[\|z\|^2 / 2]$ and $D(\mu_y \| \mu_v) = \mathbb{E}[\|\hat{z}\|^2 / 2]$.  \hspace*{\fill}$\blacksquare$
	\end{pf}	
	
	\vspace{-1em}
	
	\noindent \hyperref[prop42]{Proposition~4.2} suggests that when calculating $I(y_0^t; x_0, \break w_0^t)$ or $\bar{I}(y; x_0, w)$ by \hyperref[thm41]{Theorem 4.1} or Duncan's theorem, we actually lift the calculation and analysis of total information (rate) to a continuous function space, since the causal MMSE in \eqref{thm41_eq1} can be equivalently expressed as a divergence rate difference w.r.t. the Wiener measure on infinite-dimensional spaces. 
	
	\vspace{-0.5em}
	
	We now consider the filtering trade-off properties of $\bar{I}(y; x_0,  w)$ and $\bar{I}(y; x_0)$. Let the filtering capacity $\mathcal{C}_f := \sup_{(x_0, w, z)} \bar{I}(y; x_0, w)$ be the supremum of total information rate over all admissible pairs of $(x_0, w_0^t)$ and $z_0^t$ in \eqref{Filter_Model}. When the average power of the noise-free output is finite, i.e., $\mathbb{E}[z^\top_\tau z_\tau]  \leq \rho(\tau)$, $\forall\tau \in [0, t]$, we define the channel capacity $\mathcal{C} := \sup_{\mathbb{E}[z^\top_\tau z_\tau] \leq {\rm \rho(\tau)}}\bar{I}(y;z)$ as the supremum of input-output mutual information rate. The following theorem then reveals some fundamental relationships among the capacities, information rates, and divergence rates in the measurement channel \eqref{Filter_Model}. 
	\begin{thm}\label{thm43}
			For the filtering systems described by \eqref{Filter_Model} and \hyperref[fig2]{Fig. 2}, the total information rate $\bar{I}(y; x_0, w)$ is bounded by
			\begin{equation}\label{cor43_eq1}
				\mathcal{C}_f \geq \bar{I}(y; x_0, w) \geq \bar{I}(y; x_0 ),
			\end{equation}
			where $\bar{I}(y; x_0, w) = \bar{I}(y; x_0 )$ if $w_0^t$ and $y_0^t$ are conditionally independent given $x_0$, i.e., $w_0^t-x_0-y_0^t$ forms a Markov chain. Meanwhile, when the noise-free output is subject to the average power constraint, $\mathbb{E}[z^\top_\tau z_\tau] \leq \rho(\tau)$, $\forall\tau \in [0, t]$, we have
			\begin{equation}\label{cor43_eq2}
				\mathcal{C}_f \geq \bar{I}(y; x_0, w) + \bar{D}(\mu_y \| \mu_v) \geq \bar{I}(y; x_0) + \bar{D}(\mu_y \| \mu_v),
			\end{equation}
			where $\mathcal{C}_f = \mathcal{C} = (2t)^{-1}  \int_{0}^{t}\rho(\tau) d\tau$, and divergence rate $\bar{D}(\mu_y \| \mu_v) := \lim_{t \rightarrow \infty} D(\mu_y \| \mu_v) / t$. 
	\end{thm}
	
	 \vspace{-1.5em}
	 
	 \begin{pf}
	 	The first inequality in \eqref{cor43_eq1} follows from the definition of filtering capacity. By applying the chain rule of mutual information to $I(y_0^t; x_0, w_0^t)$ and using the non-negativity of mutual information, we can derive the second inequality in \eqref{cor43_eq1}. Meanwhile, by the chain rule, we also have $I(y_0^t; x_0, w_0^t) = I(y_0^t; x_0)$ when $w_0^t$ and $y_0^t$ are conditionally independent given $x_0$.  
	 	
	 	\vspace{-0.5em}
	 	
	 	To prove \eqref{cor43_eq2}, we first impose an average power constraint on the noise-free output such that $\mathbb{E}[z_\tau^\top z_\tau] \leq \rho(\tau)$, $\forall \tau \in [0, t]$ or $(1/t) \int_{0}^{t}\mathbb{E}[z^\top_\tau z_\tau] d\tau \leq \rho$. Since $D(\mu_{y|(x_0, w)} \| \mu_v) = \mathbb{E}[\|z\|^2 / 2]$ by \hyperref[prop42]{Proposition 4.2}, the filtering and channel capacities of \eqref{Filter_Model} then satisfy (Theorem 6.4.1 of \citealp{Ihara_1993})
	 	\begin{equation}\label{cor42_eq3}
	 		\mathcal{C}_f = \frac{\int_{0}^{t}\rho(\tau) d\tau}{2t} \geq \lim_{t \rightarrow \infty}\frac{\mathbb{E}[\|z\|^2]}{2t} = \bar{D}(\mu_{y|(x_0, w)} \| \mu_v),
	 	\end{equation}
	 	where $\mathcal{C}_f = \mathcal{C}$; the inequality is from the average power constraint and Fubini's theorem, and the divergence rate $\bar{D}(\mu_{y|(x_0, w)} \| \mu_v):= \lim_{t \rightarrow \infty} {D}(\mu_{y|(x_0, w)} \| \mu_v) / t$. Dividing both sides of \eqref{prop42_eq2} by $t$, taking the limit as $t\rightarrow \infty$, and plugging \eqref{cor42_eq3} into the results give the first inequality in \eqref{cor43_eq2}. The second inequality in \eqref{cor43_eq2} then follows \eqref{cor43_eq1}. \hspace*{\fill}$\blacksquare$	
	 \end{pf} 
	 
	 \vspace{-1.5em}
	 
	 \noindent An information-theoretic implication from \hyperref[thm43]{Theorem 4.3} is that for the measurement channel in \eqref{Filter_Model}, the channel (or filtering) capacity $\mathcal{C}$ (or $\mathcal{C}_f$), determined by the plant $\mathcal{P}$, must be large enough to offset the cost of the transmitting message, quantified by $\bar{I}(y; x_0, w)$ or $\bar{I}(y; x_0)$, and the cost of attenuating channel noise, measured by $\bar{D}(\mu_y \| \mu_v)$. Meanwhile, the capacity can be achieved when $\bar{D}(\mu_y \| \mu_v) = 0$ or $D(\mu_y \| \mu_v) = o(t)$ as $t\rightarrow \infty$.
	 
	 \vspace{-0.75em}
	 
	 Compared with the control interpretations in \hyperref[sec31]{Section 3.1}, the total information rate $\bar{I}(y; x_0, w)$ and mutual information rate  $\bar{I}(y; x_0)$ in \hyperref[thm43]{Theorem 4.3} have a more explicit and consistent filtering-theoretic explanation across different filtering systems. As \hyperref[thm41]{Theorem 4.1} shows, $\bar{I}(y; x_0, w)$ is half of the time-averaged or steady-state MS estimation error, and $\bar{I}(y; x_0) = \lim_{\varepsilon \rightarrow 0}\bar{I}(y; x_0, w)$ is half of the lowest achievable MS estimation error in the absence of process noise. When the covariance of process noise or the intensity of corresponding white noise tends to zero, $\Sigma_{\bar{w}} = \varepsilon I \geq 0$ with $\varepsilon \rightarrow 0$, we have
	 \begin{equation*}%\label{rem45_eq1}
	 	\bar{I}(y; x_0, w) \geq \lim_{\varepsilon \rightarrow 0}\bar{I}(y; x_0, w)  = \bar{I}(y; x_0),
	 \end{equation*}
	 which can be verified by \hyperref[prop42]{Proposition 4.2}, \hyperref[thm43]{Theorem 4.3}, and the fact that $\lim_{\varepsilon \rightarrow 0} I(y_0^t; w_0^t|x_0) = 0$ as the lumped process input $w_0^t \rightarrow 0$ and becomes deterministic. In practice, when the process noise exists and the filtering process $\mathcal{F}$ is not optimal, the practical MS estimation error must be larger than the minimum errors gauged by $\bar{I}(y; x_0, w)$ and $\bar{I}(y; x_0)$. Hence, a filtering-theoretic disclosure of \hyperref[thm43]{Theorem 4.3} is that to effectively transmit a message or distinguish the noise-free output from the measured output in \eqref{Filter_Model}, the channel capacity $\mathcal{C}$ (or the power of noise-free output) must be larger than $\bar{I}(y; x_0, w)$ and $\bar{I}(y; x_0)$ (or the minimum MS estimation error). In the following subsections, the lowest achievable MS estimation errors, or $\bar{I}(y; x_0, w)$ and $\bar{I}(y; x_0)$ in \hyperref[thm43]{Theorem 4.3}, will be respectively investigated in the filtering systems subject to LTI, LTV, and nonlinear plants. 
	 
	 \vspace{-0.2em}
	 
	 \begin{rem}
	 	\hyperref[thm41]{Theorems 4.1}, \ref{thm43} and \hyperref[prop42]{Proposition 4.2} apply to all filtering problems depicted by \eqref{Filter_Model} and \hyperref[fig2]{Fig. 2}, regardless of the linearity of the plant, and the stationarity or distributions of the signals and initial states. In the meantime, when the process noise $w_0^t$ is excluded or completely vanishes, total information rate $\bar{I}(y; x_0, w)$ coincides with $\bar{I}(y; x_0)$, and \hyperref[thm41]{Theorem 4.1} and \hyperref[prop42]{Proposition 4.2} remain valid, while the second inequalities in \eqref{cor43_eq1} and \eqref{cor43_eq2} reduce to equalities. 
	 \end{rem}

	\vspace{-0.5em}

	\subsection{Filtering Limits in LTI Systems}\label{sec42}
	
	\vspace{-0.5em}	
	
	To capture the filtering limitations of LTI plants, we consider the following plant and observation models
	\begin{equation}\label{Filter_LTI_Plant}
		\begin{split}
			dx_t & = Ax_t dt + Bdw_t\\
			dy_t & = Cx_t dt + \Sigma_v^{1/2} dv_t,
		\end{split}
	\end{equation}	
	\noindent where $x_t$ is the vector of the hidden states; $y_t$ is observable; $A, B$ and $C$ are time-invariant matrices of proper dimensions; $w_t$ is the lumped process noise such that $dw_t = \Sigma_{\bar{w}}^{1/2} d\bar{w}_t$; independent standard Brownian motions $\bar{w}_t$ and $v_t$ respectively denote the process and measurement noise with covariance functions $\Sigma_{\bar{w}}$ and $\Sigma_v$, and the noise-free output $z_t = Cx_t$. Let $\hat{x}_t$ and $\hat{z}_t = C\hat{x}_t$ be the estimates of states $x_t$ and output $z_t$, and the estimation error $\tilde{z}_t = z_t - \hat{z}_t = C(x_t - \hat{x}_t)$. The following optimal linear or Kalman-Bucy filter then gives an optimal MS estimation of $x_t$ and $z_t$ by conditioning on the measured output $y_0^t$ \citep{Liptser_2001, Ni_SCL_2016}:
	\begin{equation}\label{LTI_Filter}
		d\hat{x}_t = A \hat{x}_t dt + PC^\top\Sigma_v^{-1}[dy_t - C\hat{x}_t dt],
	\end{equation}
	\noindent where the optimal estimates are given by $\hat{x}_t = \mathbb{E}[x_t|y_0^t]$ and $\hat{z}_t = \mathbb{E}[Cx_t | y_0^t]$, and the positive semi-definite matrix $P_t = \mathbb{E}[(x_t - \hat{x}_t)(x_t - \hat{x}_t)^\top]$ denotes the estimation error covariance. When the LTI system \eqref{Filter_LTI_Plant} is either i) asymptotically stable or ii) both stabilizable and detectable, given any initial condition $P(t_0) \geq 0$, the covariance matrix $P_t$ converges as $t\rightarrow\infty$, i.e., $P = \lim_{t\rightarrow\infty}P_t$, which can be solved from the ARE \citep{Sivan_1972}:
	\begin{equation}\label{LTI_ARE}
		0 = AP + PA^\top + B \Sigma_{\bar{w}} B^\top - PC^\top \Sigma_v^{-1}CP.
	\end{equation}
	
	\vspace{-1em}
	
	For the filtering systems subject to the LTI plant \eqref{Filter_LTI_Plant}, irrespective of how the filter $\mathcal{F}$ is designed, the following proposition develops an equality between the mutual information rate $\bar{I}(y; x_0)$ in \hyperref[thm43]{Theorem 4.3} and the sum of open-loop unstable poles in \eqref{Filter_LTI_Plant}. 
	
	\vspace{-0.75em}
	
	\begin{prop}\label{prop45}
		For the filtering systems subject to the stabilizable and detectable LTI plant \eqref{Filter_LTI_Plant}, as in \hyperref[fig2]{Fig. 2}, we have
		\begin{equation}\label{prop43_eq1}
			\bar{I}(y; x_0) =  \sum_i\lambda_{i}^{+}(A), 
		\end{equation}
		\noindent where $\lambda_{i}^{+}(A)$ denotes the eigenvalues of matrix $A$ with positive real parts. 
	\end{prop} 
	
	\vspace{-2em}
	
	\begin{pf}
		Equation \eqref{prop43_eq1} can be proved by applying \hyperref[thm41]{Theorem 4.1} to the measurement channel in \eqref{Filter_LTI_Plant} and resorting to the ARE of Kalman-Bucy filter \eqref{LTI_ARE}. Since the analysis of LTI systems can be treated as a special case of the LTV results to be discussed later, we postpone and combine this proof with the proof of \hyperref[prop48]{Proposition 4.8}. \hspace*{\fill}$\blacksquare$
	\end{pf}  
	
	\vspace{-2em}
	
	\noindent \hyperref[prop45]{Proposition 4.5} shows that when the plant model $\mathcal{P}$ in \hyperref[fig2]{Fig. 2} is an LTI model as \eqref{Filter_LTI_Plant}, the total information rate $\bar{I}(y; x_0, w)$ and mutual information rate $\bar{I}(y; x_0)$ are bounded below or equal to the sum of open-loop unstable poles in \eqref{Filter_LTI_Plant}. Meanwhile, along with the relationships between $\bar{I}(y; x_0, w)$, $\bar{I}(y; x_0)$ and the minimum MS estimation error, we can imply that for the LTI filtering systems, $\bar{I}(y; x_0, w)$ and $\bar{I}(y; x_0)$ serve as an information-theoretic interpretation or replacement for the lowest achievable MS estimation error (in the absence of process noise, \citealp{Braslavsky_Auto_1999}), and the MS estimation error of the filtering system \eqref{Filter_LTI_Plant} must be larger than twice the sum of unstable poles. Analogous to the trade-off properties of Bode's integral in the frequency range and the entropy cost function in the time domain, $\bar{I}(y; x_0, w)$, $\bar{I}(y; x_0)$ or the lowest achievable MS estimation error serve as a filtering trade-off in the time domain: For the LTI filtering system \eqref{Filter_LTI_Plant}, if the MS estimation error of $z$ is smaller than twice the sum of unstable poles, $\bar{I}(y; x_0, w)$ or $\bar{I}(y; x_0)$, on a time interval, the MS estimation error outside this interval must be larger, and vice versa.
	
	\vspace{-0.5em}
	
	\subsection{Filtering Limits in LTV Systems}\label{sec43}
	
	\vspace{-0.5em}
	
	We then consider the LTV filtering systems with the following plant $\mathcal{P}$ and observation model
	\begin{align}\label{LTV_Filtering}
%		\begin{split}
			dx_t & = A(t)x_t dt +B(t) dw_t,  \allowdisplaybreaks \\
			dy_t & = C(t)x_t dt + \Sigma_v^{1/2}(t) dv_t, \nonumber  
%		\end{split}
	\end{align}
	\noindent where $x_t$ denotes the hidden states; the measured output $y_t$ is observable; $A(t)$, $B(t)$ and $C(t)$ are time-varying matrix functions of proper dimensions; $w_t$ is the lumped process noise that satisfies $dw_t = \Sigma^{1/2}_{\bar{w}}(t)d\bar{w}_t$; the independent standard Brownian motions $\bar{w}_t$ and $v_t$ are respectively the process and measurement noise with covariance functions $\Sigma_{\bar{w}}(t)$ and $\Sigma_v(t)$, and the definitions of noise-free output $z_t$, estimate $\hat{z}_t$, and estimation error $\tilde{z}_t$ can be implied from \eqref{Filter_LTI_Plant}. The optimal MS estimates $\hat{x}_t = \mathbb{E}[x_t|y_0^t]$ and $\hat{z}_t = C \hat{x}_t$ can be calculated from \eqref{LTV_Filtering} by resorting to the following Kalman-Bucy filter \citep{Liptser_2001, Ni_SCL_2016}: \vspace{-0.5em}
	\begin{equation}\label{LTV_Filter}
		d\hat{x}_t = A(t) \hat{x}_t dt + P(t)C^\top(t) \Sigma_v^{-1}(t)[dy_t -C(t) \hat{x}_t dt],
	\end{equation}
	where $P(t) = \mathbb{E}[(x_t - \hat{x}_t)(x_t - \hat{x}_t)^\top] = \mathbb{E}[\tilde{x}_t \tilde{x}_t^\top] $ denotes the estimation error covariance. When the LTV system \eqref{LTV_Filtering} is either i) exponentially stable or ii) both uniformly completely stabilizable and uniformly completely reconstructible, for any initial condition $P(t_0) \geq 0$, the covariance function $P(\tau)$ converges to $P_t = \lim_{\tau \rightarrow \infty} P_\tau$, which can be solved from the following RDE \citep{Sivan_1972}: \vspace{-0.5em}
	\begin{equation*}\label{LTV_DRE}
		\dot{P}^{}_t = A^{}_t P^{}_t + P^{}_t A_t^\top + B^{}_t \Sigma^{}_w(t) B_t^\top  - P^{}_tC_t^\top\Sigma_v^{-1}(t)C^{}_t P^{}_t.
	\end{equation*}
	\noindent When the process input vanishes, i.e., $\Sigma_{\bar{w}}(t) = \varepsilon I \rightarrow 0$, the optimal filter \eqref{LTV_Filter} then gives the lowest achievable MS estimation error under vanishing process noise. Since $w_0^t$ and $v_0^t$ in \eqref{LTV_Filtering} are uncorrelated and zero-mean white Gaussian, the MS estimation error attained by \eqref{LTV_Filter} is the optimal and minimum solution for all feasible filters \citep{Simon_2006}. To reveal more details about this estimation error, we decompose the original LTV system \eqref{LTV_Filtering} into an UES part with states $x_s(t)$ and an UEA part with states $x_u(t)$ as \eqref{lem220_eq1} in \hyperref[lem39]{Lemma 3.9}. The lowest achievable MS estimation error of the transformed system satisfies the following lemma. 
	
	\vspace{-1.5em}
	
	\begin{lem}\label{lem46}
		When the LTV filtering system \eqref{LTV_Filtering} under vanishing process noise is uniformly completely stabilizable and detectable, the estimation error covariance, achieved by applying the optimal filter \eqref{LTV_Filter} to the modal-decomposed system of \eqref{LTV_Filtering}, satisfies 
		\begin{equation}\label{lem54_eq0}
			\lim_{\varepsilon \rightarrow 0} \lim_{\tau \rightarrow \infty}  \mathbb{E}\left\{  \left[  \begin{matrix}
				\tilde{x}_s(\tau)\\
				\tilde{x}_u(\tau)
			\end{matrix} \right] \left[  \begin{matrix}
				\tilde{x}_s(\tau)\\
				\tilde{x}_u(\tau)
			\end{matrix} \right]^\top  \right\} = \left[\begin{matrix}
				0 & 0\\
				0& P_u(t)
			\end{matrix}\right],
		\end{equation}
		where  $\tilde{x}_s(\tau) = x_s(\tau) - \hat{x}_s(\tau)$ and $\tilde{x}_u(\tau) = x_u(\tau) - \hat{x}_u(\tau)$ are the estimation errors, and the lowest achievable MS estimation error under vanishing noise is
		\begin{equation}\label{lem54_eq02}
			\lim_{\varepsilon \rightarrow 0} \lim_{\tau \rightarrow \infty}  \mathbb{E}\left\{  \left[  \begin{matrix}
				\tilde{x}_s(\tau)\\
				\tilde{x}_u(\tau)
			\end{matrix} \right]^\top \left[  \begin{matrix}
				\tilde{x}_s(\tau)\\
				\tilde{x}_u(\tau)
			\end{matrix} \right]  \right\} = {\rm{tr}}[P_u(t)],
		\end{equation}
		where $P_u(t)$ can be solved from the following RDE
		\begin{align}\label{lem54_eq01}
			\dot{P}_u(t) = A_u(t)P_u(t) &+ P_u(t)A_u^\top(t) \allowdisplaybreaks \\
			& - P_u(t)C_u^\top(t) C_u(t)P_u(t). \nonumber
		\end{align}
	\end{lem} 
	
	\vspace{-2.5em}
	
	\begin{pf}
	See \hyperref[appD]{Appendix D} for the proof. \hspace*{\fill}$\blacksquare$
	\end{pf}
	
	\vspace{-1.5em}
	
	\begin{rem}
		Since $P_u(t)$ is the estimation error covariance of the antistable subsystem, \hyperref[lem46]{Lemma 4.6} reveals that under the vanishing process noise, only the antistable modes contribute to the MS estimation error. An LTI version of \hyperref[lem46]{Lemma 4.6} has been given by \cite{Braslavsky_Auto_1999} and can be used to prove \hyperref[prop45]{Proposition 4.5}.
	\end{rem}
	
	\vspace{-0.5em}
	
	Applying \hyperref[thm41]{Theorem 4.1} and \hyperref[lem46]{Lemma 4.6} to the LTV filtering system \eqref{LTV_Filtering} under vanishing process noise, we can obtain the following proposition that reveals some equality and inequality relationships for the total information rate or minimum MS estimation error in \hyperref[thm43]{Theorem 4.3}, regardless of how the filtering process $\mathcal{F}$ is designed. 
	
	\vspace{-0.5em}
	
	\begin{prop}\label{prop48}
		For the filtering system described by the LTV plant \eqref{LTV_Filtering} under vanishing process noise, as in \hyperref[fig2]{Fig. 2}, the mutual information rate $\bar{I}(y; x_0)$ satisfies
		\begin{align}\label{prop410_eq1}
			\hspace{-10pt} \bar{I}(y; x_0) & =\lim_{t \rightarrow \infty} \frac{1}{t} \int_{0}^{t} {\rm{tr}}[A_u(\tau)] d\tau \\
			&\hspace{30pt}- \lim_{t \rightarrow \infty} \frac{1}{2t} \int_{0}^{t} {\rm{tr}}[\dot{P}_u(\tau) P_u^{-1}(\tau)] d\tau  \allowdisplaybreaks \nonumber \\
			& \geq \sum_{i=1}^{l}n_i \underline{\lambda}_i - \limsup_{t \rightarrow \infty} \frac{1}{2}{\rm{tr}}[\dot{P}_u(t) P_u^{-1}(t)] \nonumber \allowdisplaybreaks \\
			{\rm or \hspace{5pt} } & \leq  \sum_{i=1}^{l}n_i \overline{\lambda}_i - \liminf_{t \rightarrow \infty} \frac{1}{2}{\rm{tr}}[\dot{P}_u(t) P_u^{-1}(t)], \nonumber  
		\end{align} 
		where $A_u(t)$ is the antistable part of $A(t)$; $\underline{\lambda}_i$ and $\overline{\lambda}_i$, $i =1, \cdots, l$, are the spectral values of $A_u(t)$, and the estimation error covariance matrix $P_u(t)$ satisfies \eqref{lem54_eq01}. 	
	\end{prop}	
	
	\vspace{-1.5em}
	
	\begin{pf}
		To restore the proof for \hyperref[prop45]{Proposition 4.5}, we only need to replace the following time-varying terms from \eqref{LTV_Filtering} by the time-invariant terms in \eqref{Filter_LTI_Plant}. When the lumped process noise vanishes or the covariance function $\Sigma_{\bar{w}} = \varepsilon I \rightarrow 0$, by \hyperref[lem46]{Lemma 4.6}, the lowest achievable MS estimation error of \eqref{LTV_Filtering} satisfies		
		
		\vspace{-0.5em}
		
		\begin{align}\label{prop56_eq2}
			\lim_{\varepsilon \rightarrow 0}	\lim_{\tau\rightarrow \infty} \mathbb{E}[\tilde{z}_\tau^\top \tilde{z}_\tau]  & \neweq{(a)} \lim_{\varepsilon \rightarrow 0} \lim_{\tau \rightarrow \infty}  \mathbb{E}\left\{  \left[  \begin{matrix}
				\tilde{z}_s(\tau)\\
				\tilde{z}_u(\tau)
			\end{matrix} \right]^\top \left[  \begin{matrix}
				\tilde{z}_s(\tau)\\
				\tilde{z}_u(\tau)
			\end{matrix} \right]  \right\}    \allowdisplaybreaks \nonumber\\ 
			&  \neweq{(b)} \textrm{tr}[P_u(t)C^\top_u(t)C_u(t)] \allowdisplaybreaks \nonumber\\
			&  \neweq{(c)} {\rm{tr}}[P^{1/2}_u(t)C_u^\top(t)C_u(t)P^{1/2}_u(t)] \allowdisplaybreaks \nonumber\\
			&  \neweq{(d)} {\rm{tr}}(\varPsi_t) \allowdisplaybreaks,
		\end{align}
		\noindent where (a) follows the identity $\tilde{z}_\tau = C_\tau \tilde{x}_\tau = [C_s(\tau) \  C_u(\tau)] \cdot  [\tilde{x}^\top_s(\tau) \ \tilde{x}^\top_u(\tau)]^\top$ from \eqref{lem220_eq1}; when the LTV system \eqref{LTV_Filtering} is uniformly completely stabilizable and detectable, (b) follows from \hyperref[lem46]{Lemma~4.6}, $\tilde{z}_u(t) = C_u(t) \tilde{x}_u(t)$, $P_u(t) = \mathbb{E}[ \tilde{x}_u(t) \tilde{x}_u^\top(t)]$, and the trace property $\textrm{tr}(MN) = \textrm{tr}(NM)$ for matrices $M$ and $N$ of proper dimensions; (c) utilizes the positive definiteness of covariance $P_u(t)$ and the same trace property in (b), and $\varPsi_t = P^{-1/2}_u(t)  A_u(t)P_u^{1/2}(t) + P_u^{1/2}(t)  A^\top_u(t)  P_u^{-1/2}(t) -\dot{P}_u(t) P_u^{-1}(t)$ in (d), which can be derived from \eqref{lem54_eq01}.
		
		\vspace{-0.5em}
		
		By applying \hyperref[thm41]{Theorem 4.1} to the measurement channel \eqref{LTV_Filtering} under vanishing process noise, we can derive \eqref{prop410_eq1} and the following relationships for $\lim_{\varepsilon \rightarrow 0}\bar{I}(y; x_0, w) = \bar{I}(y; x_0)$
		
		\vspace{-2em}
		
		\begin{align}\label{prop48_eq3}
			\lim_{\varepsilon \rightarrow 0}\bar{I}(y; x_0, w) & \neweq{(a)} \lim_{\varepsilon\rightarrow 0} \lim_{t \rightarrow \infty}  \frac{1}{2t} \int_{0}^{t} \mathbb{E}[\tilde{z}_\tau^\top \tilde{z}_\tau] d\tau \allowdisplaybreaks \\
			& \hspace{-15pt} \neweq{(b)} \lim_{t\rightarrow\infty} \frac{1}{2t} \int_{0}^{t} \textrm{tr}(\varPsi_\tau) d\tau\allowdisplaybreaks  \nonumber\\
			& \hspace{-15pt} \neweq{(c)} \lim_{t \rightarrow \infty} \frac{1}{2t} \int_{0}^{t} {\rm{tr}}[2A_u(\tau) - \dot{P}_u(\tau) P_u^{-1}(\tau)] d\tau \allowdisplaybreaks \nonumber\\
			& \hspace{-15pt} \newgeq{(d)} \sum_{i=1}^{l}n_i \underline{\lambda}_i - \limsup_{t\rightarrow\infty} \frac{1}{2}{\rm{tr}}[\dot{P}_u(t) P_u^{-1}(t)] \allowdisplaybreaks \nonumber\\
			{\rm or \hspace{30pt}}	& \hspace{-15pt} \newleq{(e)} \sum_{i=1}^{l}n_i \overline{\lambda}_i - \liminf_{t\rightarrow\infty} \frac{1}{2}{\rm{tr}}[\dot{P}_u(t) P_u^{-1}(t)], \nonumber
		\end{align}
		
		\vspace{-1em}
		
		\noindent where (a) utilizes \hyperref[thm41]{Theorem 4.1} and Fubini's theorem to exchange the order of integration and expectation; (b) owes to \eqref{prop56_eq2}; (c) implies the equality in \eqref{prop410_eq1}, and when \eqref{LTV_Filtering} fulfills (\hyperref[ass5]{A5})-(\hyperref[ass7]{A7}), (d) and (e) follow \hyperref[lem310]{Lemma 3.10} and L'H\^opital's rule. To restore the proof for \hyperref[prop45]{Proposition 4.5}, we notice that in the LTI scenario, the spectral values in \eqref{prop48_eq3} coincide with the real parts of the open-loop unstable poles in \eqref{Filter_LTI_Plant}, i.e., ${\rm tr}[A_u(t)] = \sum_{i} \lambda_i^{+}(A)$, and $\lim_{t \rightarrow \infty} {\rm tr}[\dot{{P}}^{}_u(t){P}_u^{-1}(t)] = 0$ as the RDE \eqref{lem54_eq01} degenerates to a steady-state ARE when \eqref{Filter_LTI_Plant} is stabilizable and detectable \citep{Kailath_TAC_1976, Braslavsky_Auto_1999}. \hspace*{\fill}$\blacksquare$ 	
		\end{pf} 
		
		\vspace{-1.5em}
		
		\hyperref[prop48]{Proposition 4.8} provides an equality and a lower bound for $\bar{I}(y; x_0, w)$ and $\bar{I}(y; x_0)$ in the LTV filtering \eqref{LTV_Filtering} regardless of the filter $\mathcal{F}$ design, and an upper bound for the (worst) MS estimation error of the Kalman-Bucy filter \eqref{LTV_Filter} under vanishing process noise. Combining \hyperref[prop48]{Proposition 4.8} with \hyperref[thm43]{Theorem 4.3}, we can imply that for the filtering systems subject to the LTV plant \eqref{LTV_Filtering}, the total information rate $\lim_{\varepsilon \rightarrow 0}\bar{I}(y; x_0, w)$ or mutual information rate $\bar{I}(y; x_0)$ serves a similar role as (or equals half) the lowest achievable MS estimation error $\lim_{\varepsilon \rightarrow 0}\mathbb{E}[\|z-\hat{z}\|^2]$ for all feasible filters in the presence of process noise. This LTV filtering limit, according to \eqref{prop410_eq1}, is determined by the spectral values of the antistable component $A_u(t)$ and the steady-state error covariance $P_u(t)$ of the optimal filter \eqref{LTV_Filter}, and shares the same time-domain properties as the LTI trade-off: When the estimation error is smaller than this filtering limit on a time interval, the error outside this interval must be larger, and vice versa. Meanwhile, the LTV result and derivation presented in \hyperref[prop48]{Proposition 4.8} cover the LTI result in \hyperref[prop45]{Proposition 4.5} as a special case, which can be explained by the same arguments in \hyperref[rem313]{Remark 3.13}.

		\subsection{Filtering Limits in Nonlinear Systems}\label{sec44}
		In the end, we consider the nonlinear filtering problem described as follows
		\begin{equation}\label{filter_model}
			\begin{split}
				dx_t & = \bar{f}(t, x_t) dt + \bar{b}(t, x_t)d\bar{w}_t\\
				dy_t & = z(t, x_t) dt + \Sigma_v^{1/2}(t) dv_t,
			\end{split}
		\end{equation}
		\noindent where $x_t$ and $y_t$ are the hidden and observable states; $\bar{f}(t, x_t) = f(t, x_t) + b(t, x_t)u(t, x_t)$ is the deterministic drift term with passive dynamics $f(t, x_t)$, control matrix $b(t, x_t)$, and control signal $u(t, x_t)$; $\bar{b}(t, x_t) = b(t, x_t) \cdot \Sigma^{1/2}_{\bar{w}}(t)$ denotes the lumped control matrix, and the process noise $\bar{w}_t$ and measurement noise $v_t$ are two independent standard Brownian motions with covariance functions $\Sigma_{\bar{w}}$ and $\Sigma_v = I$. The noise-free output $z(t, w_0^t, x_0)$ in \eqref{Filter_Model} is rewritten as the $z_x(t, x_t)$ or $z(t, x_t)$ in \eqref{filter_model}, which satisfies $z_t = z(t, w_0^t, x_0) = z_x(t, x_t) = z(t, x_t)$ and implies that $I(y_0^t; x_0, w_0^t) = I(y_0^t; x_0^t)$ by applying \hyperref[thm41]{Theorem 4.1} to \eqref{Filter_Model} and \eqref{filter_model}, respectively. For conciseness, only the SISO measurement channel with $z_t$ and $y_t \in\mathbb{R}$ is considered, and the MIMO extension has been discussed in \hyperref[rem315]{Remark 3.15}.
		
		\vspace{-0.5em}
		
		To calculate the total information rate $\bar{I}(y; x_0, w)$, or equivalently $\bar{I}(z\rightarrow y)$ and $\bar{I}(z; y)$, from \eqref{filter_model}, we resort to the identity $I(y_0^t; x_0, w_0^t) = \mathbb{E}[\|z - \hat{z}\|^2 / 2]$ in \hyperref[thm41]{Theorem 4.1}. However, neither the noise-free output $z(t, x_t)$, nor the optimal estimate $\hat{z}(t, x_t)$ can be directly measured in \eqref{filter_model}. A feasible approach to derive $\mathbb{E}[\|z - \hat{z}\|^2]$ is to adopt the following identities (\citealp{Liptser_2001}, Chapter 10 \& 13)
		\begin{equation}\label{sec53_eq1}
			\begin{split}
				\mathbb{E}[(z - \hat{z})^2] &= \mathbb{E}[(z - \hat{z})^2| y_0^\tau]\\
				& = \mathbb{E}[z^2(\tau, x_\tau)|y_0^\tau] - (\mathbb{E}[z(\tau, x_\tau)|y_0^\tau])^2,
			\end{split}
		\end{equation}
		where both $\mathbb{E}[z(\tau, x_\tau)|y_0^\tau]$ and $\mathbb{E}[z^2(\tau, x_\tau)|y_0^\tau]$ can be estimated from solving the nonlinear filtering problem \eqref{filter_model}. Meanwhile, additional regularity and differentiability constraints are required to guarantee the existence and uniqueness of the solution to \eqref{filter_model} and the pathwise uniqueness of the filter equation. Let $k(t, x)$ denote any of the functions $\bar{f}(t, x_t)$, $\bar{b}(t, x_t)$, and $z_x(t, x_0^t)$ in \eqref{filter_model}. We require $k(t, x)$ be Lipschitz such that $|k(t, x_1) - k(t, x_2)|^2 \leq K_1(|x_1 - x_2|^2)$ for some constant $K_1 > 0$, and $k^2(t, x) \leq K_2(1 + x^2)$ for some constant $K_2 > 0$. The estimated function $g(t, x_t)$, which denotes any of the functions $z(t, x_t)$ and $z^2(t, x_t)$ in \eqref{sec53_eq1}, is required be at least once differentiable at $t$ and twice differentiable at $x_t$ such that $\sup_{\tau \leq t}\mathbb{E}[g^2(\tau, x_\tau)] < \infty$, $\int_{0}^{t}\mathbb{E}[ g'_x(\tau, x_\tau)^2] d\tau < \infty$, and $\int_{0}^{t}\mathbb{E}[\mathcal{L}g(\tau, x_\tau)]^2 d\tau < \infty$, where $\mathcal{L}g(\tau, x_\tau) = g'_\tau(\tau, x_\tau)  + \break g'_{x_\tau}(\tau, x_\tau)\bar{f}(\tau, x_\tau) + (1/2)  g''_{x_\tau x_\tau}(\tau, x_\tau)  \bar{b}^2(t, x_t)$. When \eqref{filter_model} fulfills the preceding regularity and differentiability conditions, by resorting to \hyperref[thm41]{Theorem 4.1} and the Stratonovich-Kushner equation, we can calculate the total information rate $\bar{I}(y; x_0, w)$ by the following proposition. 
		
		\vspace{-0.5em}
		
		\begin{prop}\label{prop49}
				For the nonlinear filtering systems described by \eqref{filter_model}, as \hyperref[fig2]{Fig 2} shows, the total information rate $\bar{I}(y; x_0, w)$ satisfies
				\begin{align}\label{prop59_eq1}
					\hspace{-5pt} \bar{I}(y; x_0, w) 	&= \lim_{t\rightarrow \infty}  \frac{1}{2t} \int_{0}^{t} \big[ \mathbb{E}[z^2(\tau, x_\tau)|y_0^\tau] \\
					& \hspace{80pt}- (\mathbb{E}[z(\tau, x_\tau)|y_0^\tau])^2  \big] d\tau, \nonumber 
				\end{align}
				where $\pi_\tau(g) = \mathbb{E}[g(\tau, x_\tau)|y_0^\tau]$ satisfies
				\begin{align}\label{lem58_eq1}
					\pi_\tau(g) = \pi_0(g) & + \int_{0}^{\tau}\pi_s(\mathcal{L}g) ds \\
					&\hspace{-20pt} + \int_{0}^{\tau}[ \pi_s(g'_{x_s}) + \pi_s(z \cdot g) - \pi_s(z)\pi_s(g) ] d\tilde{w}_s; \nonumber
				\end{align}
				$g(\tau, x_\tau)$ denotes any of the functions $z(\tau, x_\tau)$ and $z^2(\tau, x_\tau)$ in \eqref{prop59_eq1}; $\mathcal{L}g(s, x_s) = g'_s(s, x_s) + g'_{x_s}(t, x)  \bar{f}(s, x_s) + g''_{x_s x_s}(s, x_s)  \bar{b}^2(s, x_s)$, and $\tilde{w}_\tau = \int_{0}^{\tau} dy_s - \int_{0}^{\tau} \pi_s(z) ds$ is a Wiener process w.r.t. $\mathcal{F}_\tau^y$, $\tau \in [0, t]$.
		\end{prop}
		
		\vspace{-1.5em}
		
		\begin{pf}
			By applying \hyperref[thm41]{Theorem 4.1} to the measurement channel in \eqref{filter_model}, we have
			\begin{align*}
				\bar{I}(y; x_0, w) & \neweq{(a)} \lim_{t\rightarrow \infty} \frac{1}{2t}  \mathbb{E}[\|z - \hat{z}\|^2] \allowdisplaybreaks\\
				&  \neweq{(b)}  \lim_{t\rightarrow \infty} \frac{1}{2t} \int_{0}^{t} \mathbb{E}[(z(\tau, x_\tau) - \hat{z}(\tau, x_\tau))^2]  d\tau\allowdisplaybreaks \nonumber \\
				& \neweq{(c)} \lim_{t\rightarrow \infty}  \frac{1}{2t} \int_{0}^{t} \big[ \mathbb{E}[z^2(\tau, x_\tau)|y_0^\tau] \allowdisplaybreaks \nonumber\\
				& \hspace{90pt}- (\mathbb{E}[z(\tau, x_\tau)|y_0^\tau])^2  \big] d\tau, \nonumber 
			\end{align*}
			\noindent where (a) is obtained by applying \eqref{thm41_eq1} to the measurement channel in \eqref{filter_model}, dividing the result by $t$, and taking the limit as $t\rightarrow\infty$; (b) invokes the Fubini's theorem to swap the order of integration and expectation; (c) is the application of \eqref{sec53_eq1}, and the conditional expectations $\pi_\tau(z) = \mathbb{E}[z(\tau, x_\tau)|y_0^\tau]$ and $\pi_\tau(z^2) = \mathbb{E}[z^2(\tau, x_\tau) | y_0^\tau]$ need to be estimated from solving the nonlinear filtering problem \eqref{filter_model}. By applying the one-dimensional Stratonovich-Kushner equation (\citealp{Liptser_2001}, Theorem 8.3) to \eqref{filter_model}, we obtain the optimal estimate \eqref{lem58_eq1}.  \hspace*{\fill}$\blacksquare$
	\end{pf}
	
	\vspace{-1.5em}
	
	\noindent \hyperref[prop49]{Proposition 4.9} provides a direct approach to calculate the total information rate $\bar{I}(y; x_0, w)$ in the nonlinear filtering system \eqref{filter_model}. By letting the covariance function $\Sigma_{\bar{w}} = \varepsilon I \rightarrow 0$ or $\bar{b}(t, x_t) \rightarrow 0$ in \eqref{filter_model}, \hyperref[prop49]{Proposition 4.9} can also be used to calculate the mutual information rate $\bar{I}(y; x_0)$ or $\lim_{\varepsilon\rightarrow 0}\bar{I}(y; x_0, w)$ when the process noise in \eqref{filter_model} vanishes. Consistent with the interpretations of filtering trade-off metrics in general and linear filtering systems, for nonlinear filtering system \eqref{filter_model}, the total information rate $\bar{I}(y; x_0, w)$ serves a similar role as (or equals half) the MMSE in the presence of process noise, and the mutual information rate $\bar{I}(y; x_0)$ or $\lim_{\varepsilon\rightarrow 0}\bar{I}(y; x_0, w)$ equals half the lowest achievable MS estimation error of \eqref{filter_model} in the absence of process noise, irrespective of the filter design. Meanwhile, the filtering limit and optimal estimates in \hyperref[prop49]{Proposition 4.9} cover the LTI and LTV results in \hyperref[prop45]{Propositions 4.5} and \ref{prop48} as special cases, as we can easily restore the linear optimal or Kalman-Bucy filter when applying \eqref{lem58_eq1} to a linear filtering system \eqref{Filter_LTI_Plant} or \eqref{LTV_Filtering}.
	
	\vspace{-1em}

	\begin{ack}                              
		This work was supported by AFOSR, NASA and NSF. In this preprint, the authors also thank the readers and staff at arXiv.org.
	\end{ack}

	\def\thesection{\Alph{section}}
	\setcounter{section}{1}
	\section*{Appendix A: Control Trade-off in a Scalar System}\label{appA}
	To give a concrete example on the control trade-off and analysis method introduced in \hyperref[sec31]{Section 3.1}, we consider the following open-loop scalar system
	\begin{equation}\label{app_eq1}
		dy_t = \alpha \cdot y_t dt + de_t,
	\end{equation}
	where $\alpha > 0$ implies that \eqref{app_eq1} possesses an open-loop unstable pole, and $\alpha \leq 0$ stands for a stable pole. To stabilize the closed-loop system, as \hyperref[fig1]{Fig. 1} shows, we adopt the control mapping 
	\begin{equation}\label{app_eq2}
		u_t = -k \cdot y_t,
	\end{equation}
	where  $k > 0$ is the control gain. The lumped control mapping or control channel \eqref{Gen_Ctrl_Err} then satisfies
	\begin{equation}\label{app_eq3}
		de_t = u_t dt + dw_t = -k \cdot y_t dt +  dw_t,
	\end{equation}
	where $w_t$ is a standard Brownian motion representing noise. By substituting \eqref{app_eq2} and \eqref{app_eq3} into \eqref{app_eq1}, we can then use the following Ornstein-Uhlenbeck process to describe the output signal $y_0^t$ 
	\begin{equation}\label{app_eq4}
		dy_t = (\alpha - k) \cdot y_t dt + dw_t = \bar{\alpha} \cdot y_t dt + dw_t,
	\end{equation}
	which has a solution
	\begin{equation}\label{app_eq5}
		y_t = y_0 \cdot \exp(\bar{\alpha} t) + \int_{0}^{t} \exp\left[ \bar{\alpha} (t - s)      \right] dw_s .
	\end{equation}
	When the closed-loop pole $\bar{\alpha} = \alpha - k \leq 0$, the closed-loop system \eqref{app_eq4} is stable, and the solution \eqref{app_eq5} implies a convergent or stable output signal.

	\hyperref[thm31]{Theorem 3.1} provides two alternative approaches to calculate $I(e_0^t; x_0) = ({1}/{2}) \cdot ( \mathbb{E}[\|u\|^2] - \mathbb{E}[\|\hat{u}\|^2]) = ({1} / {2}) \cdot \mathbb{E}[\|u - \hat{u}\|^2]$, which require the explicit expressions for $\mathbb{E}[\|u\|^2]$, $\mathbb{E}[\|\hat{u}\|^2]$, and $\mathbb{E}[\|u - \hat{u}\|^2]$. Assuming that the initial condition $y_0$ in \eqref{app_eq5} is normal $y_0\sim \mathcal{N}(0, \sigma_y^2)$ and substituting \eqref{app_eq5} into \eqref{app_eq2}, we have
	\begin{align*}
		\mathbb{E}[\|u\|^2] & =  \mathbb{E}\Big[ \int_{0}^{t}u^2(\tau) d\tau  \Big]  \allowdisplaybreaks \\
		& = k^2 \cdot \Big[ \exp(2\bar{\alpha} \cdot t) \cdot \sigma_y^2 -  \frac{1}{{2\bar\alpha}} \left( 1 -  \exp(2 \bar{\alpha} \cdot t)  \right) \Big].
	\end{align*}
	By applying the Kalman-Bucy filter (\citealp{Liptser_2001}, Theorem~10.3) to the filtering problem described by \eqref{app_eq3} and \eqref{app_eq4} or $du_t = \bar{\alpha} \cdot u_t dt - k \cdot dw_t$, we can then calculate $\mathbb{E}[\|\hat{u}\|^2]$ by solving the following filter equation
	\begin{equation*}
		d \hat u_t = \bar \alpha \cdot  \hat u_t dt +[-k+p(t)] \cdot [de_t - \hat u_t dt],
	\end{equation*} 
	or $\mathbb{E}[\|u - \hat{u}\|^2]$ by solving the following error covariance equation
	\begin{equation}\label{app_eq7}
		\dot p(t) =  -p^2(t)+2\alpha \cdot p(t), 
	\end{equation}  
	where $\hat{u}_t = \mathbb{E}[u_t|e_0^t] = -k \cdot \hat{y}_t = -k \cdot \mathbb{E}[y_t|e_0^t]$, and $p(t) = \mathbb{E}[(u(t) - \hat{u}(t))^2]$ is the estimation error covariance. Since the solution to \eqref{app_eq7} is more explicit, we use $I(e_0^t; x_0) =  \mathbb{E}[\|u - \hat{u}\|^2 / 2]$ to compute $I(e_0^t; x_0)$ and $\bar{I}(e; x_0)$. The solution to \eqref{app_eq7} is
	\begin{equation}\label{app_eq8}
		p(t) = \frac{ 2 \alpha \cdot  p(0) \cdot \exp(2\alpha t)}{p(0) \cdot \exp(2\alpha t)-p(0)+2\alpha}
	\end{equation}
	where, by letting $t\rightarrow \infty$, we compute the asymptotic error covariance $\lim_{t\rightarrow \infty}p(t) = 2 \alpha$ when $\alpha > 0$, and  $\lim_{t\rightarrow \infty}p(t) = 0$ when $\alpha \leq 0$. With the estimation error covariance given in \eqref{app_eq8}, we compute the mutual information $I(e_0^t; x_0)$ in (\ref{app_eq1}-\ref{app_eq3}) via
	\begin{align*}
		I(e_0^t; x_0) & = \frac{1}{2}  \mathbb{E}[\|u - \hat{u}\|^2]= \frac{1}{2}  \int_0^t p(\tau) \ d\tau \allowdisplaybreaks \\
		& \hspace{-5pt} = \frac{1}{2} \big [\log \left(p(0)\exp(2\alpha t)- p(0) +2\alpha  \right )- \log(2\alpha ) \big], \nonumber
	\end{align*}
	where the Fubini's theorem is used to swap the order of integration and expectation. Consequently, the mutual information rate 
	\begin{equation}\label{app_eq9}
		\bar{I}(e; x_0) = \lim_{t \rightarrow \infty} \frac{1}{2t} \int_0^t p(\tau) \ d\tau =\begin{cases}
			\alpha \hspace{10pt} {\rm if \ } \alpha > 0, \\
			0 \hspace{11.5pt} {\rm if \  } \alpha \leq 0, 
		\end{cases}  
	\end{equation}
	where the limit is evaluated by using the L'H\^opital's rule and the limit of \eqref{app_eq8}. Since $\bar{I}(e; x_0)$ equals the (sum of) open-loop unstable poles, as \eqref{app_eq9} shows, we can imply that the mutual information rate $\bar{I}(e; x_0)$ is identical to the famous Bode's integral \citep{Freudenberg_TAC_1985}, which serves as a control trade-off characterizing the disturbance rejection ability of scalar control system (\ref{app_eq1})-(\ref{app_eq3}). Meanwhile, with \hyperref[prop32]{Proposition 3.2} and preceding results, we can also calculate $D(\mu_{e|x_0} \| \mu_w) =  \mathbb{E}[\|u\|^2 / 2]$ and $D(\mu_{e}\|\mu_w)  = \mathbb{E}[\|\hat{u}\|^2 / 2]= I(e_0^t; x_0) - \mathbb{E}[\|u\|^2 / 2]$. Rigorous analysis on the control trade-offs $I(e_0^t; x_0)$ and $\bar{I}(e; x_0)$ is performed in \hyperref[sec3]{Section 3}, and more examples in LTI, LTV and nonlinear control systems can be found by mimicking the steps in this example.

	\section*{Appendix B: Stability and Sensitivity of LTV Systems}\label{AppB}	\vspace{-0.5em}
	We start with the following $n$-dimensional homogeneous LTV system
	\begin{equation}\label{eq_a1}
		\dot{x}_t = A(t) x_t.
	\end{equation}
	Let $\{x^{(i)}(t), 1 \leq i \leq n \}$ be $n$ linearly independent solutions to~\eqref{eq_a1}. We then define the \textit{fundamental solution matrix} $X(t) = [x^{(1)}(t),  \cdots, x^{(n)}(t) ]$, which satisfies $\dot{X}(t) = A(t)X(t)$, and the state transition matrix $\Phi_A(t, t_0) = X(t)X^{-1}(t_0)$. When $\{p_i\}_{i=1}^n$ is an orthonormal basis in $\mathbb{R}^n$, the \textit{characteristic numbers} are defined by 
	\begin{equation*}
		\lambda_i(p_i) = \limsup_{t\rightarrow\infty}\frac{1}{t} \log \|X(t)p_i \|.
	\end{equation*}
	When the sum of the characteristics numbers $\sum_{i=1}^{n} \lambda_i(p_i)$ is minimized, the orthonormal basis $\{p_i\}_{i=1}^n$ is called normal, and the $\lambda_i$'s are called the \textit{Lyapunov exponents}. The characteristic numbers and Lyapunov exponents in LTV systems serve the same role as the real parts of the eigenvalues of $A$ in LTI systems. Definitions of \textit{uniform exponential stability} and \textit{antistability} are given as follows.
	\begin{defn}
		The matrix function $A(t)$ is uniformly exponentially stable (UES) if there exist positive constants $\gamma$ and $\delta$ such that
		\begin{equation*}  
			\|\varPhi_A(t, \tau)\| \leq \gamma e^{-\delta(t - \tau)}
		\end{equation*}
		for all $t \geq \tau$.
	\end{defn}
	
	\begin{defn}
		The matrix function $A(t)$ is uniformly exponentially antistable (UEA), if there exist positive constants $\gamma$ and $\delta$ such that
		\begin{equation*}  
			\|\varPhi_A(t, \tau)\| \leq \gamma e^{-\delta(t - \tau)}
		\end{equation*}
		for all $t \leq \tau$.
	\end{defn}
	The LTV system~\eqref{eq_a1} is said to possess an \textit{exponential dichotomy}, if there exists a projection $P$, and positive constants $\gamma$ and $\delta$ such that
	\begin{equation*}
		\begin{split}
			\|X(t)PX(\tau)^{-1}\| &\leq \gamma e^{-\delta(t - \tau)}, \qquad \forall \ t\geq \tau,\\
			\|X(t)(I-P)X(\tau)^{-1}\| &\leq \gamma e^{-\delta(t- \tau)}, \qquad \forall \ t\leq \tau.\\
		\end{split}
	\end{equation*}
	If $\textrm{rank}(P) = n_s$, exponential dichotomy implies that $n_s$ fundamental solutions are UES, whereas $n_u = n - n_s$ are UEA. The \textit{dichotomy spectrum} of LTV system \eqref{eq_a1} is the set of real values $\lambda$ for which the translated systems $\dot{x}(t) = [A(t) - \lambda I]  x(t)$ fail to have a dichotomy. In general, the dichotomy spectrum is a collection of compact non-overlapping intervals 
	\begin{equation*}
		\mathcal{S}_{\textrm{dich}} = \cup_{i=1}^m[\underline{\lambda}_i, \overline{\lambda}_i],
	\end{equation*}
	where $m \leq n$ and $\underline{\lambda}_1 \leq \overline{\lambda}_1 < \underline{\lambda}_2 \leq \overline{\lambda}_2 < \cdots < \underline{\lambda}_m \leq \overline{\lambda}_m$. It is straightforward to verify that when $\lambda_0 < \underline{\lambda}_1$ all trajectories of $\dot{x}(t) = [A(t) - \lambda_0 I]  x(t)$ are unbounded, and when $\lambda_{m+1} > \overline{\lambda}_m$ all trajectories of $\dot{x}(t) = [A(t) - \lambda_{m+1} I] x(t)$ are bounded. When~\eqref{eq_a1} is an LTI system or the matrix function $A(t)$ in~\eqref{eq_a1} is periodic, each of these intervals $[\underline{\lambda}_i, \overline{\lambda}_i]$ is a point equivalent to a Lyapunov exponent $\lambda_i$. The dichotomy is then known as a \textit{point spectrum}, and the system~\eqref{eq_a1} is said to be \textit{regular}.

	For the LTV system $\Sigma_L$ defined in~\eqref{LTV_Plant}, its sensitivity operator $S$ (or sensitivity system $\Sigma_S$) has the following state-space representation 
	\begin{equation*}
		\Sigma_S = \left[\begin{array}{c;{3pt/2pt}c}
			A(t)-B(t)C(t) & B(t)\\ \hdashline[3pt/2pt]
			-C(t) & I 
		\end{array}\right],
	\end{equation*}
	which serves the same role as the (state-space representation of) \textit{sensitivity function} $S(s) = 1 / [1 + L(s)]$ in LTI systems~\citep{Wu_TAC_1992}. When $A(t)$ admits an exponential dichotomy and the pair $(A(t), B(t))$ is stabilizable, the sensitivity operator has an inner/outer factorization $S = S_i S_o$, where the state-space representations of these two factors are respectively
	\begin{equation*}
		\Sigma_{S_i} = \left[\begin{array}{c;{3pt/2pt}c}
			A(t)-B(t)B(t)^\top X(t) & B(t)\\ \hdashline[3pt/2pt]
			-B(t)^\top X(t) & I 
		\end{array}\right],
	\end{equation*}
	and
	\begin{equation*}
		\Sigma_{S_o} = \left[\begin{array}{c;{3pt/2pt}c}
			A(t)-B(t)C(t) & B(t)\\ \hdashline[3pt/2pt]
			B(t)^\top X(t) -C(t) & I 
		\end{array}\right].
	\end{equation*}
	Define the kernel of the outer factor as
	\begin{equation*}
		\hat{S}_o(t, \tau) = [B(t)^\top X(t) - C(t)]\Phi_{A-BC}(t, \tau) B(\tau),
	\end{equation*}
	for all $t \geq \tau$, and then the following integral
	\begin{equation*}
		\mathcal{B} := \lim_{t\rightarrow\infty} \frac{1}{4t}\int_{-t}^{t}\textrm{tr}[\hat{S}_o(\tau, \tau)] d\tau
	\end{equation*}
	is an analogue of Bode's sensitivity integral for continuous-time LTV systems~\citep{Wu_TAC_1992, Iglesias_LAA_2002}. More comprehensive introductions on the stability and sensitivity of LTV systems refer to~\cite{Coppel_1978}, \cite{Dieci_JNA_1997}, \cite{Iglesias_LAA_2002}, \cite{Tranninger_CSL_2020} and references therein.

	\section*{Appedix C: Proof of Corollary 3.12}\label{AppC}	\vspace{-0.5em}
	The first inequality in \eqref{prop316_eq1} can be directly obtained by applying \hyperref[thm34]{Theorem 3.4} to the control channel that consists of the LTV plant \eqref{LTV_Plant} and nonlinear control mapping \eqref{Gen_Ctrl_Ctrl}. To prove \hyperref[cor312]{Corollary 3.12}, it remains to verify that $\bar{I}(e; x_0) \geq  \lim_{t\rightarrow \infty} t^{-1} \int_{0}^{t}{\rm tr}[A_u(\tau)]d\tau  \geq \sum_{i=1}^{l} n_i \underline{\lambda}$. Since the matrix function $A(t)$ in \eqref{LTV_Plant} possesses an exponential dichotomy, we can decompose the LTV system \eqref{LTV_Plant} into UES and UEA parts as \hyperref[lem39]{Lemma 3.9} shows, in which the UEA part can be described by
	\begin{equation}\label{prop38_eq2}
		d{x}_u(t) = A_u(t)x_u(t) dt + B_u(t) de(t),
	\end{equation}
	where $x_u(t)$ denotes the antistable modes. The solution to \eqref{prop38_eq2} can be expressed as \citep{Platen_2006}
	\begin{align}\label{prop38_eq25}
		x_u(t) & = \varPhi^{}_{A^{}_u}(t, 0)  x^{}_u(0) + \int_{0}^{t} \varPhi^{}_{A_u}(t, \tau) B_u(\tau)  de(\tau) \nonumber \allowdisplaybreaks\\
		& = \varPhi^{}_{A^{}_u}(t, 0) \left[ x^{}_u(0) + \bar{x}^{}_u(t)  \right],
	\end{align}
	where $\varPhi_{A_u}(\tau, \tau_0)$ is the state transition matrix of $A_u(t)$ from $\tau_0$ to $\tau$, and $\bar{x}^{}_u(t) = \varPhi^{-1}_{A_u}(t, 0)  \int_{0}^{t} \varPhi_{A_u}(t, \tau)   B(\tau)  de(\tau)$. Since the closed-loop system is internally mean-square stable, we have
	\begin{align}\label{prop38_eq3}
		+\infty >	M >\	&\log\left[  \det\left( \mathbb{E}\left[ x_u(t) x^\top_u(t)  \right] \right)  \right] \nonumber \allowdisplaybreaks\\
		= \ & 2  \log \left[ \det \left(  \varPhi_{A_u}(t, 0) \right) \right] + \log\left[  \det\left( \Gamma  \right) \right] \nonumber \allowdisplaybreaks\\
		\neweq{(a)} \ & 2 \int_{0}^{t}\textrm{tr}[A_u(\tau)] d\tau + \log\left[  \det\left( \Gamma  \right) \right]  , \allowdisplaybreaks
	\end{align} 
	where $\Gamma = \mathbb{E}[  (x_u(0) + \bar{x}_u(t))(x_u(0) + \bar{x}_u(t))^\top ]$, and step (a) follows from the Liouville's formula. Meanwhile, the mutual information $I(e_0^t; x_0)$ satisfies
	\begin{align}\label{prop38_eq4}
		I(e_0^t; x_0)	& \newgeq{(a)} I(e_0^t; x_u(0)) \nonumber \allowdisplaybreaks\\
		& \newgeq{(b)} I(\bar{x}_u(t); x_u(0)) \nonumber \allowdisplaybreaks\\
		& \neweq{(c)} h(x_u(0)) - h(x_u(0)|\bar{x}_u(t))  \allowdisplaybreaks\\
		& \neweq{(d)} h(x_u(0)) - h(x_u(0) + \bar{x}_u(t)|\bar{x}_u(t))\nonumber \allowdisplaybreaks\\
		& \newgeq{(e)} h(x_u(0)) - h(x_u(0) + \bar{x}_u(t))\nonumber \allowdisplaybreaks\\
		&   \newgeq{(f)} h(x_u(0)) - \frac{1}{2} \log(2\pi\textrm{e})^k - \frac{1}{2} \log\left(  \det \left(  \Gamma \right)  \right),\nonumber
	\end{align}
	where (a) follows the data processing inequality, $I(x; y|z) \geq I(x; g(y)|z)$, as $x_u(0)$ is a function of $x_0$ according to \eqref{LTV_Plant} and \eqref{lem220_eq1}; (b) also follows the data processing inequality as $\bar{x}_u(t)$ is a function of $e_0^t$ in \eqref{prop38_eq25}; (c) uses the identity $I(x; y) = h(y) - h(y|x)$; (d) employs the invariance of entropy, $h(x|z) = h(x + g(z)|z)$ where $g(\cdot)$ is a function; (e) adopts the fact that $I(x; y) = h(y) - h(y|x) \geq 0$, and (f) uses the maximum entropy, $h(x) \leq h(\bar{x}) = (1/2) \cdot \log[(2\pi e)^n \det(\Sigma_x)]$ where $\bar{x}$ is a Gaussian vector with the same covariance as $x$. Hence, substituting \eqref{prop38_eq3} or $-\log(\det(\Gamma) ) > -M + 2\int_{0}^{t}\textrm{tr}(A_u(\tau)) d\tau$ into \eqref{prop38_eq4} yields
	\begin{align}\label{prop38_eq5}
		I(e_0^t; x_0) \geq {h(x_u(0))} & - \frac{1}{2}\log(2\pi\textrm{e})^k \allowdisplaybreaks \\
		& - \frac{M}{2} + \int_{0}^{t}\textrm{tr}[A_u(\tau)] d\tau. \nonumber
	\end{align}	
	Dividing both sides of \eqref{prop38_eq5} by $t$ and taking the limit as $t\rightarrow \infty$, we have
	\begin{align*}
		\bar{I}(e; x_0)  & \geq \lim_{t\rightarrow \infty} \bigg[ \frac{h(x_u(0))}{t}  - \frac{1}{2t} \log(2\pi e)^n - \frac{M}{2t}  \allowdisplaybreaks \\
		& \hspace{110pt} + \frac{1}{t} \int_{0}^{t}{\rm tr}[A_u(\tau)]d\tau \bigg] \allowdisplaybreaks \\
		& = \lim_{t\rightarrow \infty} \frac{1}{t} \int_{0}^{t}{\rm tr}[A_u(\tau)]d\tau \geq \sum_{i=1}^{l} n_i \underline{\lambda}_i  \allowdisplaybreaks,
	\end{align*}
	where the last inequality follows \hyperref[lem310]{Lemma 3.10}. This completes the proof.

	\section*{Appendix D: Proof of Lemma 4.6}\label{appD}\vspace{-0.5em}
	When the matrix function $A(t)$ in \eqref{LTV_Filtering} admits an exponential dichotomy, by \hyperref[lem39]{Lemma 3.9}, we can decompose \eqref{LTV_Filtering} into the following modal form 
	\begin{align}\label{lem54_eq1}
		\left[\begin{matrix}
			dx_s(t)\\
			dx_u(t)
		\end{matrix}\right]
		& = 
		\left[\begin{matrix}
			A_s(t) & 0\\
			0 & A_u(t)
		\end{matrix}\right]\left[\begin{matrix}
			x_s(t)\\
			x_u(t)
		\end{matrix}\right]dt + \left[
		\begin{matrix}
			B_s(t)\\
			B_u(t)
		\end{matrix}\right]dw(t) \allowdisplaybreaks \nonumber\\
		dy(t) & = \left[\begin{matrix}
			C_s(t) & C_u(t)
		\end{matrix}\right]\left[\begin{matrix}
			x_s(t) \\
			x_u(t)
		\end{matrix}\right] dt + dv(t).
	\end{align}
	Let $\bar{x}(t) = [x_s^\top(t), x_u^\top(t)]^\top$ be the state of modal-decomposed systems, and $\tilde{\bar{x}}(t) = [x_s^\top(t) - \hat{x}_s^\top(t), x_u^\top(t) - \hat{x}_u^\top(t)]^\top$ be the estimation error. When \eqref{lem220_eq1} or \eqref{LTV_Filtering} is uniformly completely stabilizable and detectable, we have $\lim_{\tau \rightarrow \infty}\mathbb{E}[\tilde{\bar{x}}(\tau) \tilde{\bar{x}}^\top(\tau)] = P_t$ and $\lim_{\tau \rightarrow \infty}\mathbb{E}[\tilde{\bar{x}}^\top(\tau) \tilde{\bar{x}}(\tau)] = \textrm{tr}(P_t)$, where $P_t$ can be solved from the following RDE
	\begin{align}\label{lem54_eq2}
%		\begin{split}
			&\dot{P}_t =	 \left[\begin{matrix}
				A_s(t) & 0\\
				0 & A_u(t)
			\end{matrix}\right]P_t 
			+ \varepsilon^2 \left[\begin{matrix}
				B^{}_s(t)B^\top_s(t) & B^{}_s(t)B_u^\top(t)\\
				B^{}_u(t)B^\top_s(t) & B^{}_u(t)B^\top_u(t)
			\end{matrix}\right] \nonumber \\
			&+ P_t\left[\begin{matrix}
				A^\top _s(t) & 0\\
				0 & A^\top _u(t)
			\end{matrix}\right]
			- P_t\left[\begin{matrix}
				C_s^\top(t) C_u(t) & C_s^\top(t) C_u(t)\\
				C_u^\top(t) C_s(t) & C_u^\top(t) C_u(t)
			\end{matrix}\right] P_t.
%		\end{split}
	\end{align}
	For the singularly perturbed equation \eqref{lem54_eq2}, we seek its solution $P_t = P^\top_t$ in the form of power series in $\varepsilon$ \citep{Jameson_AMO_1975, Naidu_1988, Braslavsky_Auto_1999}:
	\begin{equation}\label{lem54_eq3}
		P(t) = \left[\begin{matrix}
			\varepsilon^2 P_s(t) + O(\varepsilon^3) & \varepsilon^2 P^{}_0(t) + O(\varepsilon^3)\\
			* & P_u(t) + O(\varepsilon)
		\end{matrix}\right].
	\end{equation}
	Substituting the leading terms of \eqref{lem54_eq2} into the RDE \eqref{lem54_eq2}, $P_s(t)$, $P_0(t)$ and $P_u(t)$ in \eqref{lem54_eq3} then satisfy
	\begin{subequations}\label{lem54_eq4}
		\begin{align}
			&\varepsilon^2 \dot{P}_s(t) = \varepsilon^2[ A_s(t)P_s(t)  + P_s(t) A^\top_s(t) \allowdisplaybreaks \\
			&\hspace{102pt} + B_s(t) B^\top_s(t) ] + O(\varepsilon^4) \nonumber \allowdisplaybreaks\\
			&\varepsilon^2 \dot{P}_0(t) = \varepsilon^2  [ A_s(t)P_0(t) + P_0(t)A_u^\top(t) + B_s(t)B_u^\top(t) \nonumber \allowdisplaybreaks \\
			&\hspace{84pt} - P_s(t)C_s^\top(t)C_u(t)P_u(t) \allowdisplaybreaks \\
			&\hspace{84pt} - P_0(t)C_u^\top(t)C_u(t)P_u(t)] + O(\varepsilon^4) \nonumber \allowdisplaybreaks \\
			&\dot{P}_u(t) = A_u(t)P_u(t) + P_u(t)A_u^\top(t) \allowdisplaybreaks  \\
			& \hspace{84pt} - P_u(t)C_u^\top(t) C_u(t)P_u(t) + O(\varepsilon^2) \nonumber
		\end{align}
	\end{subequations}
	After setting $\varepsilon = 0$ in \eqref{lem54_eq3} and \eqref{lem54_eq4}, we find that the only nonzero element in $P(t)$ is $P_u(t)$, as \eqref{lem54_eq0} shows, and $P_u(t)$ is the positive definite solution of \eqref{lem54_eq01}. This completes the proof.

	{\small
		\bibliography{myref}           % and a bib file to produce the                                         
	}

	\end{document}